\documentclass[a4paper,11pt]{article}
\usepackage{jheppub} 
\usepackage{lineno}
\usepackage{amsmath}
\usepackage{tikz}
\usepackage{spectralsequences}
\usepackage{hyperref}
\usepackage{jheppub}
\usepackage{tikz-cd}
\usepackage{pgfplots}
\usepackage{tikz-3dplot}
\usepackage{braket}
\usepackage{tikz,lipsum,lmodern}
\usepackage[most]{tcolorbox}

\usetikzlibrary{matrix}

\title{\boldmath Mixed Anomalies of Magnetic Symmetries}

\author{Aiden Sheckler}
\affiliation{Department of Physics, University of California, San Diego,\\
9500 Gilman Drive, La Jolla CA 92093-0319, USA}

\emailAdd{asheckler@ucsd.edu}

\abstract{We resolve the existence of mixed 't Hooft anomalies between the electric and magnetic (solitonic) symmetries in $\sigma$-models and gauge theories. We identify the anomaly as naturally originating from a higher group in the Whitehead tower of the target space. In particular, the magnetic charges (topological solitons) can be interpreted as higher connections on the stages of the Whitehead tower, when pulled back to spacetime. This allows us to derive the form of the $(d+1)$-dim topological theory which classifies the anomaly by inflow. We also find novel features of these anomalies resulting from the mapping class group of the target space. We give many explicit examples, including a discussion of the cases where the higher magnetic symmetries become non-invertible. }

\begin{document}

\maketitle
\flushbottom 

\section{Introduction}
\label{sec1}

Symmetries are central to our investigation of quantum field theory. A fundamental property of a global symmetry is its ability to possess a 't Hooft anomaly. This structure is most commonly identified as an obstruction to promoting the global symmetry to a gauge symmetry. These anomalies are extraordinarily useful pieces of information about the theory. They are topological in nature, and so must be matched across renormalization group flows \cite{tHooft:1979rat}. This provides useful constraints for studying theories in the IR when we can not track the RG flow directly. The diagnosis of a ‘t Hooft anomaly, as well as an understanding of its origin, is therefore a powerful tool in the analysis of a theory. \\

There is a particular type of ‘t Hooft anomaly which arises repeatedly across many different quantum field theories. It appears as a \textit{mixed} anomaly between two distinct types of internal symmetries, which means it is an obstruction to gauging both symmetries simultaneously. The two types of symmetries here are often referred to as “electric” and “magnetic” symmetries. Electric symmetries are manifest symmetries of the fields, such as flavor symmetries which appear as transformations of the field space. Magnetic symmetries are often more subtle, and result from nontrivial topology of the field space. These names are a reference to particularly well-known examples of these anomalies in 4d abelian gauge theories \cite{Gaiotto:2014kfa}, where the electric and magnetic symmetries are 1-form symmetries generated by the field strength and its dual. However these sorts of anomalies are much more ubiquitous, and are indeed quite common among quantum field theories. Developing an understanding of the origin of these anomalies is the central goal of this paper. \\

A common class of quantum field theories are $\sigma$-models, where the fields are scalars $\sigma_i$ which parametrize some target space $X$. The global electric and magnetic symmetries in this case have a clear distinction. The electric symmetries are geometric symmetries of the target space. They are implemented as isometry transformations of the field variables $\sigma_i$ which leave the action invariant. Magnetic symmetries result from nontrivial topology of the target space, and are often referred to as “solitonic symmetries”. This is because their charged objects are “solitons”: defects of the theory around which the fields $\sigma_i$ can wind into topologically nontrivial configurations \cite{RevModPhys.51.591}. These include familiar objects such as domain walls, vortices, and skyrmions.  \\

The fact that the charged objects of magnetic symmetries are extended operators fits in nicely to the modern understanding of \textit{generalized symmetries} \cite{Gaiotto:2014kfa, Cordova:2022ruw, McGreevy:2022oyu, Brennan:2023mmt, Bhardwaj:2023kri, Costa:2024wks}. In the past decade our understanding of symmetry in quantum field theory has experienced a rapid evolution, yielding a new wealth of structures that we can interpret as symmetries. Among these are so-called \textit{higher-form symmetries}, whose charged objects are extended operators. The magnetic symmetries we discuss here are comfortably at home in this language, and can often be interpreted as prime examples of higher-form symmetries. Indeed, in a $d$-dim quantum field theory, a nontrivial homotopy group of the target space $\pi_k(X) $ indicates the existence of a $(d-k-1)$-form symmetry, which we denote $M_k$. The charged objects of this symmetry are solitons which are supported on $(d-k-1)$-dim submanifolds of spacetime. \\

Mixed anomalies between the electric and magnetic symmetries of $\sigma$-models are extremely common. Indeed, it was recently conjectured in \cite{Pace:2023kyi} that such a mixed anomaly \textit{always} exists, as long as both symmetries are present. Among other evidence, this was motivated by observations that anomalies may be diagnosed by studying the charges of the magnetic symmetry defect operators \cite{Gaiotto:2014kfa, Bhardwaj:2022dyt}. However gauge theories also possess mixed anomalies between electric and magnetic symmetries, where the electric symmetries now also manifest as higher-form symmetries, and the magnetic charges are topologically nontrivial configurations of the gauge field, such as 't Hooft lines or instantons. It is therefore natural to extend this conjecture to include gauge theories. We review many examples of these anomalies in \hyperref[sec22]{Sec 2.2}, both in the context of sigma models and gauge theories. \\ 

The purpose of this work is to resolve this conjecture by identifying precisely when such an anomaly exists. We will not only diagnose the conditions which give rise to such an anomaly, but we will also be able to completely characterize it as long as the symmetries remain invertible. The modern characterization of an anomaly of a $d$-dim quantum field theory is by \textit{anomaly inflow}, which specifies a bulk $(d+1)$-dim topological quantum field theory known as the \textit{anomaly theory}.  The original anomalous theory can then be thought of as living on the $d$-dim boundary of the anomaly theory. In this work, we will find the precise conditions under which the anomaly exists, and will derive the form of the bulk anomaly theory which characterizes it. The data we use will be only the symmetries and topology of the target space $X$ (and its generalization in the context of gauge theories). \\

The key to our argument will be to look at a topological structure associated to the target space $X$, known as its \textit{Whitehead tower}. This is defined as a series of topological spaces which form a tower of “covering” fibrations over $X$. The different stages are known as \textit{Whitehead stages}, and their role in life is to “kill” progressively higher-homotopy data of $X$. In this sense, they are the higher-homotopy generalizations of the universal covering space of $X$, which kills only $\pi_1$. Our central insight is that the charges of the magnetic symmetry (the topological solitons) can be interpreted topologically as defining higher-connections on the fibrations for these Whitehead stages, when pulled back over spacetime. \\

The reason this interpretation turns out to be so powerful is because of the indirect approach we take to study the mixed electric-magnetic anomalies. Instead of studying the magnetic symmetry directly, we will first study the topological relationship between the electric symmetry and the magnetic \textit{charges}.  The motivation for this approach will be described more thoroughly in \hyperref[sec31]{Sec 3.1}, but there is strong historical precedent for it, as there is a long history of understanding the topological relationship between anomalous symmetries and the corresponding symmetry charges \cite{Tachikawa:2017gyf, Wang:2017loc, Kapustin:2014lwa, Kapustin:2014zva, Cordova:2018cvg, Bhardwaj:2017xup,Carqueville:2016kdq, Bhardwaj:2023wzd, Bhardwaj:2023ayw, Gaiotto:2017yup, Barkeshli:2014cna, Kapustin:2013uxa, Frohlich:2009gb, Bartsch:2022mpm, Bhardwaj:2022kot}. We will then see that this relationship naturally predicts the existence of a mixed anomaly between the electric and magnetic symmetries. By quantifying this relationship as a higher group extension \cite{Cordova:2018cvg,Benini:2018reh}, we will also be able to derive the topological class of the anomaly theory in one higher dimension which characterizes the anomaly by inflow. \\

Using this approach, we will also propose a novel feature of the anomaly which may result from a nontrivial mapping class group of the target space. The mapping class group of a topological space describes the group of connected components of the isometries, such as orientation reversal. The mapping class group also has a well-known action on the homotopy groups of a topological space, which is quantified by the \textit{Torelli group} of the space. We will find that such a nontrivial action will have the effect of twisting the product of the electric and magnetic symmetries, as well as twisting the mixed anomaly between them. \\

The story becomes more interesting when we consider multiple magnetic symmetries in the same theory, so that we have multiple nontrivial homotopy groups $\pi_k(X)$ leading to symmetries of different degrees. It was shown recently \cite{Chen:2022cyw,Chen:2023czk, Pace:2023mdo} that these symmetries can combine nontrivially, and that generically the higher-form symmetries can become non-invertible. This means that instead of being described by a group-like structure, the symmetry defect operators which implement the symmetry form a higher fusion category. There has been much work and interest in non-invertible symmetries in recent years (see \cite{Schafer-Nameki:2023jdn, Shao:2023gho} for a review), as well as their anomalies \cite{Kaidi:2023maf, Apruzzi:2023uma, Bartsch:2023wvv, Choi:2023xjw, Zhang:2023wlu, Cordova:2023bja, Anber:2023pny, Antinucci:2023ezl, Seifnashri:2024dsd}. Our arguments above for describing the electric-magnetic anomalies apply when the magnetic symmetries remain invertible, but the approach for describing the anomaly theory breaks down when the symmetry is non-invertible. Nevertheless we believe our techniques are still capable of detecting the presence and some features of the anomaly, even when the higher magnetic symmetries are non-invertible. We give a discussion regarding these points in \hyperref[sec5]{Section 5}. \\

 A useful description of the symmetries of a theory are encoded in the Symmetry Topological Field Theory (SymTFT), which exists in one higher dimennsion \cite{Freed:2022qnc, Kapustin:2014gua, Ji:2019jhk, Apruzzi:2021nmk, Bhardwaj:2023ayw, Gaiotto:2020iye}. In \cite{Pace:2023mdo} it was shown that the SymTFT for the magnetic symmetries is Morita equivalent to that of a symmetry given by the higher-group which appears in the Postnikov tower of the target space. It was then proposed that if one could also incorporate the electric symmetry into the SymTFT then an extension with this higher group would indicate a mixed anomaly. While we do not use the SymTFT formalism in this work, our results in some sense give a realization of this expectation, as we find a natural extension between the electric symmetry and the magnetic \textit{charges}. \\

 The plan for the paper is as follows. In \hyperref[sec2]{Section 2} we review the electric and magnetic symmetries of bosonic sigma models and gauge theories, which allows us to frame the main conjecture that there always exists an anomaly between these symmetries. We review many known examples of these mixed anomalies. In \hyperref[sec3]{Section 3} we give the main argument of this work which shows the existence of the mixed anomaly between the electric and magnetic symmetry. We begin by reviewing motivation for our approach of studying the anomaly by the magnetic charges. We then review the idea of the Whitehead tower of the target space $X$, and show that the magnetic charges can be interpreted as higher connections on the stages of the Whitehead tower. More mathematical background and elaboration on this argument is provided in the \hyperref[appa]{Appendix}. This allows us to understand the topological relationship between the electric symmetry action and the magnetic charges as a higher group. We show that this higher group structure leads to the existence of a mixed anomaly between the electric and magnetic symmetries, and allows us to derive a formula for the corresponding anomaly theory. We then consider the effect of the mapping class group of the target space, and show that it can twist both the symmetry and the anomaly. In \hyperref[sec4]{Section 4} we revisit our examples and show that they can be understood from the perspective introduced in \hyperref[sec3]{Section 3}. This allows us to rederive the anomalies in each example using our anomaly formula and show that they match. In \hyperref[sec5]{Section 5} we discuss the case of magnetic symmetries associated to higher homotopy groups. We begin by giving a review of recent work which shows that the magnetic symmetries associated to higher homotopy groups can be non-invertible, and present examples, including several novel examples such as the $SO(3)$ $\sigma$-model and $SO(3)$ gauge theory. We then argue that our methods can be extended to characterize the anomalies associated to higher magnetic symmetry as long as the symmetry remains invertible. \\ 

\newpage

\section{Solitonic Symmetries and Their Anomalies}
\label{sec2}

We begin by reviewing the two types of symmetries we are interested in: (i) electric symmetries which are isometries of the target space, and (ii) magnetic symmetries which come from the topology of the target space, and whose charges are solitonic defects. This allows us to frame the question of primary interest to us: when does there exist a mixed anomaly between these two kinds of symmetries? We give a number of examples of theories which contain both such symmetries, all of which possess such anomalies. 

\subsection{Symmetries of $\sigma$-Models}
\label{sec21}

The models of primary interest to us are $\sigma$-models in some $d$-dim spacetime $M_d$. They are described by massless scalar fields $\{\sigma_i\} : M_d \to X $ that parametrize some target manifold $X$. The symmetries of such models typically fall into two types:\\

\noindent \textbf{Electric symmetries:} These are the geometric symmetries of the target space itself. The target space $X$ often takes the form of a symmetric space, with a group of isometries $G$ which act on $X$. An important example is the case when our $\sigma$-model is the low-energy effective theory for some spontaneous symmetry breaking $H \to K$, so that $X = H/K$ is a quotient space. In this case, the isometries $G$ descend from automorphisms of the original symmetry $G\subset \text{Aut}(H)$. \\

The action of the $G$-symmetry on the fields is equivalent to the representation of the $G$-action on the coordinates of $X$. That is, given $g \in G$ interpreted as a map $g : X \to X$ then the fields transform as $\vec \sigma \to g\circ \vec \sigma $. This is clearly an ordinary invertible 0-form symmetry. \\

\noindent \textbf{Magnetic symmetries:} These are symmetries which correspond to topologically nontrivial configurations of the fields $\sigma : M_d\to X$. They result from nontrivial homotopy groups $\pi_k(X)$ of the target space $X$. The objects charged under these symmetries are solitonic defects (vortices, skyrmions, domain walls, etc.). If we denote the set of homotopy classes of $\sigma$ as a map from $M_d$ into $X$ by $[\sigma]\in [M_d, X]$, then a nontrivial class in this set indicates the presence of one or more solitons. The insertion of a solitonic defect takes us to a different topological charge sector, described by the winding of the field $\sigma$ around the defect. \\

The traditional understanding of these solitonic symmetries is that they are invertible higher-form symmetries, with a different species of soliton for each nontrivial homotopy group $\pi_k(X)$. For such a homotopy group, the associated soliton is described by a $(d-k-1)$-dim defect $A_{d-k-1}$, and the charge of the defect is equivalent to the winding of the map $[\sigma] \in \pi_k(X)$ around $A_{d-k-1}$. The defect can be defined by excising a small tubular neighborhood $U$ of $A_{d-k-1}$ with boundary $\partial U\cong A_{d-k-1} \times \mathbb{S}^k$. Then a soliton of charge $\alpha \in \pi_k(X)$ is defined by enforcing the boundary condition that $\sigma |_{\partial U} : A_{d-k-1}\times \mathbb S^k \to X$ has winding $\alpha \in [\mathbb S^k,X]\subset \pi_k(X)$. \\

Since the charged objects associated to a homotopy group $\pi_k(X)$ are $(d-k-1)$-dim defects, we expect that the associated solitonic symmetry is an invertible $(d-k-1)$-form symmetry, which we will denote $M_k^{(d-k-1)}$. Since the group of charges of this symmetry is $\pi_k(X)$, this means that the group describing this symmetry is the Pontryagin dual group $M_k = \hat \pi_k(X)\equiv \text{Hom}(\pi_k,U(1))$. The topological symmetry defect operators which generate symmetry transformations are topological operators $U_{\alpha}(\Sigma_k)$ supported on a $k$-dim submanifold $\Sigma_k$ for $\alpha \in \hat \pi_k(X)$. These symmetry defect operators can act on the solitonic charges $A_{d-k-1}$ of charge $\beta$ by linking $\Sigma_k$ with $A_{d-k-1}$, which yields a phase $\alpha(\beta)\in U(1)$. It is sometimes the case that we have an explicit expression for a $k$-form homotopy invariant $J^{(k)}$ constructed from our fields which can measure the winding number by being integrated over $\Sigma_k$. In the case of continuous symmetries then such invariants often correspond to de Rham cohomology classes, and can be seen to be the conserved Noether current for the solitonic symmetry $M_k$. In such cases then the symmetry defect operators can be written in an explicit form such as $U_{\theta}(\Sigma_k) = \exp(i\theta \int_{\Sigma_k} J^{(k)})$. \\

However, recently it has been realized that the above description in terms of invertible symmetries can break down in cases when we have multiple species of soliton, corresponding to multiple nontrivial homotopy groups $\pi_k(X)$ \cite{Chen:2022cyw, Chen:2023czk, Pace:2023mdo}. The intuitive reason for this is that the homotopy type of $X$ may not cleanly decompose into products of the homotopy groups $\pi_k$, and instead the different solitonic symmetries can mix nontrivially. The symmetries that would correspond to the higher homotopy groups can then become non-invertible. We review this full description of the solitonic symmetry in \hyperref[sec51]{Section 5.1}. \\

For the time being, we will therefore restrict only to the lowest nontrivial homotopy group $\pi_k(X)$, which for simplicity we assume to be abelian. In this case then the corresponding solitonic symmetry $M_k$ is guaranteed to be an invertible $(d-k-1)$-form symmetry. \\

\subsubsection*{Gauge Theories}

Beyond just $\sigma$-models, other quantum field theories can also have magnetic symmetries. A familiar setting where such symmetries arise is in gauge theories. These are theories which are equipped with a gauge field for some group $\mathcal{G}$, corresponding to a connection on a principal $\mathcal{G}$-bundle. If $\mathcal{G}$ is abelian then we can also consider higher-gauge theories, where the gauge field is a higher-form field and corresponds to a higher connection on a gerbe. In gauge theories, magnetic symmetries can result from nontrivial topology of $\mathcal G$, in particular, nontrivial homotopy groups $\pi_k(\mathcal G)$. \\

This can be understood by realizing that the principal $\mathcal{G}$-bundle can have topologically-distinct nontrivial bundle structures. These possible bundle structures are classified by homotopy classes of continuous maps $f : M_d \to B\mathcal G $ into the classifying space $B\mathcal G$, so that the set of topologically nontrivial bundles is $[M_d,B\mathcal G]$. If $\mathcal G$ has lowest nontrivial homotopy group $\pi_k(\mathcal G)$ then by the suspension isomorphism $\pi_{k+1}(B\mathcal G)\cong \pi_k(\mathcal G)$. \\

In a completely analogous way to the solitonic symmetries of a $\sigma$-model, this results in a higher-form magnetic symmetry of the gauge theory. The charged objects are $(d-k-2)$-dim defects. Examples include the 't Hooft lines of abelian gauge theory, corresponding to the $\pi_2(BU(1))\cong \pi_1(U(1))=\mathbb Z$, or the instantons of nonabelian Yang-Mills, corresponding to the $\pi_4(B\mathcal G)\cong \pi_3(\mathcal G)$ for compact semisimple Lie group $\mathcal G$. The $\pi_1(SO(n))\cong \mathbb Z_2$ also gives a magnetic symmetry in $SO$ gauge theory, which leads to distinct choices of configurations of Wilson/'t Hooft lines \cite{Aharony:2013hda, Lee:2021crt}.  

\subsubsection*{Anomalies}

Now that we have a good understanding of the two types of symmetries that arise in our theories, we can present the primary question of interest in this paper: \\

\begin{tcolorbox}[enhanced,attach boxed title to top center={yshift=-3mm,yshifttext=-1mm},
  colback=blue!5!white,colframe=blue!75!black,colbacktitle=red!80!black,
  title=Question,fonttitle=\bfseries,
  boxed title style={size=small,colframe=red!50!black} ]
  When does there exist a mixed 't Hooft anomaly between the electric symmetry $G$ and a magnetic symmetry $M_k$?  
\end{tcolorbox} 

\bigskip

In \cite{Pace:2023kyi} it was conjectured that such an anomaly always exists, which was motivated by studying the charges of the homotopy defects in a $G$-symmetric presentation of the $X$ $\sigma$-model.\\ 

In \hyperref[sec3]{Section 3} we take up the resolution to this question, and identify precisely the condition under which such an anomaly exists between $G$ and the lowest $M_k$. Moreover, we derive the anomaly theory that classifies the anomaly by inflow, and identify new general features of the anomaly associated to the mapping class group of $X$. In \hyperref[sec5]{Section 5} we discuss the extension to higher homotopy groups.  \\

As a preview, we present the answer to the above question: 

\begin{tcolorbox}[enhanced,attach boxed title to top center={yshift=-3mm,yshifttext=-1mm},
  colback=blue!5!white,colframe=blue!75!black,colbacktitle=red!80!black,
  title=Answer,fonttitle=\bfseries,
  boxed title style={size=small,colframe=red!50!black} ]
  A mixed anomaly between $G$ and $M_k$ exists if the electric symmetry $G$ participates in a nontrivial higher group extension when it is lifted to the $k^{th}$ stage $X^k$ of the Whitehead tower of $X$, characterized by a nontrivial extension class $e \in H^{k+1}(BG,\pi_k)$.  In terms of background gauge fields $A$ for $G$ and $B$ for $M_k$, the topological class of the  $(d+1)$-dim anomaly theory can be expressed as:
  $$ \alpha_{SPT}= \exp\Big(i\int_Y A^*e\cup B\Big) $$
\end{tcolorbox} 

The goal of \hyperref[sec3]{Section 3} will be to prove this result.

\subsection{Examples}
\label{sec22}

We now give several examples of theories with both electric and solitonic symmetry, and describe the mixed anomalies between them. 

\subsubsection{The Compact Boson}

The simplest nontrivial example of a theory with both electric and magnetic symmetries is the compact boson, given by a single scalar $\phi$ with target space $X=\mathbb S^1$, so that $\phi(x)+2\pi \sim  \phi(x)$. The action is simply
\begin{equation}
    S= \int d^d x \ \frac{1}{4\pi}\partial_{\mu}\phi\partial^{\mu}\phi 
\end{equation}
The electric symmetry is just the ordinary 0-form $U(1)$ translation symmetry $\phi \to \phi + a$, with Noether current $j=-\frac{i}{2\pi}d\phi$, while the magnetic symmetry is the $(d-2)$-form $U(1)^{(d-2)}$ winding symmetry with Noether current $\tilde{j}=\frac{1}{2\pi}\star d\phi$. Under dualization, the electric and magnetic symmetries are exchanged, so that the winding symmetry acts on the dual higher Maxwell field as translation by a flat connection. \\

There exists a mixed anomaly between the electric and magnetic symmetries \cite{Cheng:2022sgb}, which can be seen by coupling to background gauge fields for each of the symmetries:
\begin{equation}
    S[A_1, B_{d-1}] = \int d^dx \Big[\frac{1}{4\pi}(d\phi - A)^2+\frac{i}{2\pi}B\wedge(d\phi -A) \Big]
\end{equation}
Then under a gauge transformation $A_1 \to A_1+d\Lambda_0$ and $B_{d-1}\to B_{d-1}+d\beta_{d-2}$ the action varies as $\delta S=  - \frac{i}{2\pi}\int \beta\wedge dA$. The $(d+1)$-dim anomaly theory, living on a bulk $(d+1)$-dim manifold $Y$ with $\partial Y=X$, which cancels this anomaly by inflow is then given by:
\begin{equation} \label{CBanom}
    \alpha_{SPT} = \exp\Big(\frac{i}{2\pi}\int_Y B\wedge d A\Big)
\end{equation}

\subsubsection{The Torus}

A theory with target space the torus $X=\mathbb T^2$ is just two decoupled copies of the compact boson, so the above story follows through. We can of course generalize to the $N$-torus but we take $\mathbb T^2$ for simplicity. There is a 0-form electric $U(1)\times U(1)$ translation symmetry and a dual magnetic $(d-2)$-form $U(1)\times U(1)$ winding symmetry. Coupling to background gauge fields as before we find a mixed anomaly between these symmetries with a bulk anomaly theory:
\begin{equation}
    \alpha_{SPT} = \exp\Big(\frac{i}{2\pi}\int_Y \sum_{i=1,2}B_i\wedge d A_i\Big)
\end{equation}

However we can also take into account the fact that the torus has an additional discrete symmetry coming from its mapping class group $MCG(\mathbb T^2)=SL(2,\mathbb Z)$. This means that the full electric symmetry of the torus is really $G = U(1)^2\rtimes SL(2,\mathbb Z)$, so that when we coupled to background gauge field $\tilde A$ for $G$, we can additionally include a discrete part $\tilde A=(A_1,A_2,\alpha)$ where $A_1,A_2$ are the above gauge fields for $U(1)^2$ and $\alpha$ is a background connection on a $SL(2,\mathbb Z)$ principal bundle, which can be interpreted as the insertion of a network of codimension-1 symmetry defect operators, across which a $SL(2,\mathbb Z)$ transformation is implemented. \\

We can ask how this extra piece of the background gauge field contributes to the anomaly. We give a schematic argument for now, but the full justification will be revisited in \hyperref[sec34]{Sec 3.4} and \hyperref[sec42]{Sec 4.2}. The question we need to answer is how do we couple the discrete part of the gauge field to the magnetic background gauge field $B$, as in the above Lagrangian? The key observation is to note that the action of $SL(2,\mathbb Z)$ acts nontrivially on the homotopy charges via a representation $\rho : SL(2,\mathbb Z)\to \text{Aut}(\pi_1)=\text{Aut}(\mathbb Z\times \mathbb Z) $. Conversely, this also induces an action on the magnetic symmetry defect operators defined by the background gauge field $B$. Intuitively, this results from the fact that an action $\rho$ on $\pi_1$ induces an action $\hat \rho$ on $\hat \pi_1= \text{Hom}(\pi_1,U(1))$ defined by $\hat\rho (a)=a\circ \rho$ . This suggests that the discrete gauge field $\alpha$ can \textit{act} on the magnetic background gauge field $B$. In terms of symmetry defect operators, it means that if a symmetry defect operator $U_{g}$ corresponding to $B$ passes through a symmetry defect operator $V_a$ corresponding to $\alpha$, then $U_g$ comes out the other side as $U_{\rho_a(g)}$. \\

The lesson is that the two background gauge fields should be coupled in such a way which preserves this action. The natural candidate for the resulting anomaly theory is then the \textit{twisted} $(d+1)$-dim theory:

\begin{equation} \label{tanom}
    \tilde \alpha_{SPT} = \exp\Big(\frac{i}{2\pi}\int_Y B\cup_{\hat\rho} d A\Big)
\end{equation}

where $\cup_{\hat\rho}$ is the twisted cup product \cite{Benini:2018reh,localcoef}. We will see the general argument for this result in \hyperref[sec34]{Sec 3.4}. \\

Another novel feature of the torus is that there also exists a new conserved current $* \mathcal J = \frac{1}{4\pi^2} d\phi \wedge d\phi$ which generates a $(d-3)$-form $U(1)$ solitonic symmetry which measures Hopf-linked vortices of each species \cite{Pace:2023mdo}. \\

\subsubsection{Generalized Maxwell}

Generalized ($p$-form) Maxwell theory is free abelian gauge theory with $p$-form gauge field $a_p$, which is a $p$-connection on a higher principal $U(1)$ $p$-bundle. This is just a higher-form generalization of the compact boson, which can be seen as 0-form Maxwell theory. There is an electric $p$-form symmetry that shifts the gauge field by a flat connections $a_p\to a_p+\Lambda_p$, as well as a magnetic $(d-p-2)$-form symmetry whose conserved current is the field strength $F_{p+1/}/2\pi$. Once again, under dualization the electric and magnetic symmetries are interchanged, so that the magnetic symmetry manifests as a translation symmetry for the dual gauge field. The magnetic symmetry can be seen as a solitonic symmetry from the fact that the higher bundle structure is topologically classified by a map $f$ into the higher classifying space $f : M_d \to B^pU(1)$, and using the suspension isomorphism to see that $\pi_{p+1}(B^pU(1))\cong \mathbb Z$. \\

As with the compact boson, we can see the electric-magnetic mixed 't Hooft anomaly by coupling to background gauge fields $A_{p+1}$ for the electric symmetry and $B_{d-p-1}$ for the magnetic symmetry, and performing a gauge transformation. The bulk anomaly theory is given by: 
\begin{equation}\label{EManom}
    \alpha_{SPT} = \exp\Big(\frac{i}{2\pi}\int_Y B_{d-p-1}\wedge d A_{p+1}\Big)
\end{equation}
A central insight made in \cite{Freed:2006yc,Freed:2006ya} which will be important for our discussion is that this anomaly is equivalent to a projective representation between the electric and magnetic charges on the Hilbert space. \\

\subsubsection{The $\mathbb{CP}^1$ Model}

We can consider the 4d $\sigma$-model with target space $X = \mathbb{CP}^1\cong \mathbb S^2$. A convenient parematrization is to use $\mathbb{CP}^1\cong SU(2)/U(1)$, and thus we can describe the model using two complex scalars $\vec \Phi=(\Phi_1, \Phi_2)$ with the constraint $\vec\Phi \cdot \vec \Phi^{\dagger} = 1$ along with a $U(1)$ gauge redundancy $\vec \Phi(x) \sim e^{i\alpha(x)}\vec \Phi(x)$. This redundancy is captured by an auxiliary $U(1)$ gauge field $A = i\Phi ^{\dagger}\cdot d\Phi$ which is the canonical connection for the Hopf fibration. \\

The electric symmetries clearly come from the rotations of the sphere, and thus are given by $G= O(3)$, whose connected piece is just $SO(3)$. The lowest magnetic symmetry comes from $\pi_2(\mathbb S^2)\cong \mathbb Z$, so in 4d this is a 1-form $U(1)$ symmetry, and the charged lines are vortices. The conserved current which measures the $\pi_2$ winding around the vortices can be expressed as the curvature of the Hopf connection $\star J = dA/2\pi$. \\

There is a mixed anomaly between the magnetic $\pi_2$ vortex symmetry and the electric $SO(3)$ rotation symmetry \cite{Pace:2023kyi,Cordova:2018acb, Brennan:2023vsa, DHoker:2024vii}. This originates from the fact that since the fields $\vec \Phi$ transform in the fundamental of $SU(2)$ then the vortices can transform projectively under the global $SO(3)$ electric symmetry of $SU(2)/U(1)$, which implies they carry fractional spin-1/2 charge under $SO(3)$. This is the phenomenon of symmetry fractionalization \cite{Brennan:2022tyl, Delmastro:2022pfo}, and it is indicative of a mixed anomaly of the form:
\begin{equation}\label{cpanom}
    \alpha_{SPT} = \exp\Big(i\pi\int_{Y_5} w_2(A_1)\cup \beta B_2\Big)
\end{equation}
where $A_1, B_2$ are the background gauge fields for the electric and magnetic symmetry respectively. \\

In 4d the $\pi_3(\mathbb S^2)\cong \mathbb Z$ also gives rise to a 0-form $U(1)$ magnetic symmetry whose charged objects are ``Hopfions". Naively the current which measures the $\pi_3$ winding is given by the Hopf invariant $\frac{1}{4\pi} AdA$. However there is an obvious issue which is that this current is not a gauge invariant operator. We will return to this symmetry in \hyperref[sec5]{Section 5} when we discuss higher homotopy. \\

\subsubsection{The $SU(N)$ Model}

Another familiar example that arises in the context of chiral symmetry breaking in 4d QCD is the $\sigma$-model with target space $SU(N)$. The electric symmetry is given by the automorphisms of $SU(N)$, which are $G= PSU(N)\rtimes \mathbb Z_2$ if $N\geq 3$ or else $G = PU(2)$ for $N=2$. There is a $(d-4)$-form $U(1)_B$ magnetic baryon symmetry coming from the $\pi_3(SU(N))\cong \mathbb Z$, whose charged objects in 4d are skyrmions. \\

There is a well-known mixed anomaly in 4d between the electric and magnetic symmetry, which is required for anomaly matching of the $[SU(N)]_{L,R}^2 \times U(1)_B$ 't Hooft anomaly of QCD. The 5d anomaly theory is related by descent to the anomaly polynomial $I_6 = c_2(PSU(N)) \wedge c_1(U(1)_B)$, and thus takes the form:
\begin{equation} \label{suanom}
    \alpha_{SPT}=\exp\Big(\frac{i}{16\pi^2}\int_{Y_5}\text{Tr}(F\wedge F)\wedge B_1\Big)
\end{equation}
where $B_1$ is the background gauge field for $U(1)_B$. \\

\subsubsection{$SU(N)$ Gauge Theory}

Consider Yang-Mills theory with $SU(N)$ gauge group in 5d. This theory possesses an electric $\mathbb Z_N^{(1)}$ 1-form symmetry corresponding to the center $Z(SU(N))\cong \mathbb Z_N$, whose charged objects are Wilson lines. However there is also a magnetic $U(1)_I$ 0-form symmetry referred to as \textit{instanton symmetry}, since the charged objects under this symmetry are instantons. The associated topological current for this symmetry is just $J_I=\frac{1}{16\pi^2}\text{Tr} F\wedge F$, which measures the instanton number. \\

In \cite{BenettiGenolini:2020doj} it was shown that there exists a mixed anomaly between the electric $\mathbb{Z}_N^{(1)}$ 1-form symmetry and the magnetic 0-form $U(1)_I$ instanton symmetry. This anomaly can be seen by turning on a background 2-form $\mathbb Z_N$ gauge field $B_2$ for the electric symmetry, which corresponds to including $PSU(N)=SU(N)/\mathbb Z_N$ connections in the path integral. Since $PSU(N)$ gauge field configurations can have fractional instantons, the above instanton number is not invariant under large gauge transformations of the background field $B_2$. This signals a mixed anomaly between the electric and magnetic symmetries, which can be canceled via inflow from the 6d bulk anomaly theory:
\begin{equation}
    \alpha_{SPT}=\exp\Big(i\frac{2\pi i }{4\pi N}\int_{Y_6}dA\cup \mathcal P(B)\Big)
\end{equation}
where $A$ is the background gauge field for the 0-form $U(1)_I$ instanton symmetry and $\mathcal P(B)\in H^4(Y_6,\mathbb Z_{2N})$ is the Pontryagin square.

\newpage

\section{Anomalies from Charge Extension}
\label{sec3}

Our starting point for understanding the mixed anomalies between the electric isometry symmetry $G$ and the magnetic symmetry $M_k$ will be to understand how $G$ interacts with the \textit{charges} of the magnetic symmetry. For the time being, we will restrict ourselves to a solitonic symmetry associated only to the lowest nontrivial homotopy group $\pi_k(X)$, which we assume to be abelian. The group of homotopy charges is then precisely $\pi_k(X)$. The motivation for focusing on the charges in order to understand the anomaly has a long history, which we briefly review. 

\subsection{Extensions, Anomalies, and Charges}
\label{sec31}

The topological relationship between symmetry extensions, anomalies, and charges has been well studied \cite{Tachikawa:2017gyf, Wang:2017loc, Kapustin:2014lwa, Kapustin:2014zva, Cordova:2018cvg, Bhardwaj:2017xup,Carqueville:2016kdq, Bhardwaj:2023wzd, Bhardwaj:2023ayw, Gaiotto:2017yup, Barkeshli:2014cna, Kapustin:2013uxa, Frohlich:2009gb, Bartsch:2022mpm, Bhardwaj:2022kot}, and is still a subject of active interest. The situation is best understood in the context of abelian symmetries. \\

It was shown in \cite{Tachikawa:2017gyf} that under the gauging of a finite abelian symmetry $A$, these structures are exchanged. We can unpack this statement into several cases. First, if the original theory possesses a symmetry $A$ then the gauged theory will possess a dual $(d-2)$-form symmetry given by the Pontryagin dual $\hat A=\text{Hom}(A,U(1))$, which is the group of charges of the original symmetry. If the original symmetry $A$ participates in a central group extension with another symmetry in the theory, with extension class $e$, then after gauging $A$ the dual symmetry in the gauged theory possesses a mixed anomaly with the rest of the symmetry, where the topological class of the anomaly theory is determined by $e$. Conversely, the existence of a mixed anomaly between $A$ and the rest of the symmetry in the original theory yields a new symmetry in the gauged theory described by an extension with $\hat A$. If the anomaly takes a simple form, then the resulting extension is just a group extension, with extension class determined by the original anomaly. However if the anomaly takes a more complicated form then the resulting symmetry can be non-invertible, such as in \cite{Kaidi:2021xfk}. \\

This story extends to continuous abelian symmetries. A case of particular interest to us is generalized ($p$-form) Maxwell theory. There it can be shown in detail using the formalism of differential cohomology that the existence of the mixed anomaly between the electric $p$-form symmetry and the magnetic $(d-p-2)$-form symmetry is equivalent to the fact that the Hilbert space forms a nontrivial projective representation between the electric and magnetic charge sectors \cite{Freed:2006yc, Freed:2006ya}. This perspective was a fundamental inspiration for the approach taken in this paper. \\

Our primary interest is in regard to solitonic symmetries, whose groups of charges are the homotopy groups $\pi_k(X)$. These are all discrete abelian groups beyond the fundamental group, and their Eilenberg-Maclane spaces $B^k \pi_n(X) = K(\pi_n,k) $ carry an abelian group structure. For this reason, the above results give us strong incentive to study the interaction between the charges $\pi_k(X)$ and the electric symmetry $G$ of the theory, in order to investigate the existence of a mixed anomaly between the magnetic symmetry and the electric. We should then expect that the signature of such an anomaly might lie in an extension between the magnetic charge algebra and $G$. \\

Recent developments in our understanding of symmetries and anomalies further supports this. A popular proposal is that all information regarding the symmetries of a quantum field theory can be captured in a single higher-dimensional bulk topological quantum field theory known as the symmetry topological field theory (SymTFT) \cite{Freed:2022qnc, Kapustin:2014gua, Ji:2019jhk, Apruzzi:2021nmk, Bhardwaj:2023ayw}. This data includes the symmetry defects, charges, and anomalies. Thus the above relationships between features of the symmetry are encoded in the SymTFT and its boundary conditions.  \\

A useful feature of the SymTFT perspective is that it elegantly encapsulates the above relationships between symmetry extensions, anomalies, and charges under gauging. Indeed, given two symmetries and their associated fusion $(d-1)$-categories, if these symmetries are exchanged under gauging then their fusion categories are said to be \textit{Morita equivalent}. The SymTFTs of two Morita equivalent symmetries are equal \cite{Bhardwaj:2023ayw}. The procedure of gauging is relegated to an operation on the boundary conditions. \\

In \cite{Pace:2023kyi, Pace:2023mdo} this approach was used to study the solitonic symmetries of a $\sigma$-model, by realizing that the dual symmetry to the full magnetic symmetry $\text{Rep}^{d-1}(\mathbb G^{(d-1)})$ is equivalent to that of a higher-group gauge theory $\text{Vec}(\mathbb G^{(d-1)})$. The SymTFT of such a symmetry is much more amenable to study in comparison to the full (possibly non-invertible) higher-form magnetic symmetry $\text{Rep}^{d-1}(\mathbb G^{(d-1)})$, and so one can work with the SymTFT for  $\text{Vec}(\mathbb G^{(d-1)})$ instead, with the security that the results should be equivalent.  Moreover, the authors conjecture that if one could formulate the SymTFT for the entire symmetry of the theory, including the electric symmetry $G$, then the anomaly between $G$ and the magnetic symmetry would manifest as an extension between $G$ and this higher group of magnetic charges. \\

While we do not use the formalism of the SymTFT in this work, we will follow this idea of investigating the relationship between the magnetic charges and the electric symmetry $G$. Indeed, we will see that such an extension generically exists, and that as a consequence it determines the mixed anomaly in the invertible case. 

\subsection{Charge Extensions from the Whitehead Tower}
\label{sec32}

Now we will try to understand in what sense there is an extension between the global electric symmetry $G$ and the solitonic charges $\pi_k(X)$ of the magnetic symmetry $M_k$. To do this, we require a better understanding of what it means to insert a general configuration of solitonic charges $Q$. \\

The operation of inserting a solitonic defect $Q$ of charge $\alpha$ requires excising a submanifold $A_{d-k-1}$ from our spacetime, along which the defect is supported. The defect is then defined by the property that as we wind around a $k$-cycle in $M_d$ which links $A_{d-k-1}$ then the fields $\sigma$ are constrained to wind around $X$ by an amount specified by the soliton charge in $\alpha\in \pi_k(X)$ \footnote{Here we are abusing notation, and it should be understood that by $M_d$ we mean the spacetime $M_d \setminus A_{d-k-1} $ where the submanifold $A_{d-k-1}$ along which the soliton is supported has been excised.}. Topologically, a general soliton configuration $Q$ thus determines a map in $\text{Hom}(H_k(M_d),\pi_k)$. Since we are working locally around the defect, and don't need to incorporate possible spacetime torsion, this therefore defines a class $Q \in H^k(M_d,\pi_k)$. \\

It is useful to represent this cocycle by a map $ Q : M_d \to B^k\pi_k$ where $B^k\pi_k=K(\pi_k,k)$ is the $k$th Eilenberg-Maclane space of $\pi_k$. Here we are using the fact that any cocycle $a \in H^n(M,A)$ of a topological space $M$ can be classified by the homotopy class of a map $ a :M \to K(A,n)$. The utility of this expression for the cocycle will become clear shortly. \\

The natural setting to understand these soliton configurations $Q$ is in the $k^{th}$ stage $X^k$ of the \textit{Whitehead tower} of $X$. This space is the $k$-connected generalization of the universal cover $X^1$ of $X$. The Whitehead tower is constructed as a sequence of fibrations over $X$:
\[\begin{tikzcd}
	& \vdots \\
	{B\pi_{2}(X)} & {X^{2}} \\
	{\pi_{1}(X)} & {X^1} \\
	& X
	\arrow[from=1-2, to=2-2]
	\arrow[from=2-1, to=2-2]
	\arrow[from=2-2, to=3-2]
	\arrow[from=3-1, to=3-2]
	\arrow[from=3-2, to=4-2]
\end{tikzcd}\]
The  fibration defining the $k^{th}$ stage is $B^{k-1}\pi_k \to X^k \to X^{k-1}$. The homotopy groups of each stage satisfy $\pi_n(X^k) = 0$ for $n \leq k$ and $\pi_n(X^k)=\pi_n(X)$ for $n> k$. Each stage $X^n$ can be thought of as ``killing" the homotopy group $\pi_n$, while preserving all higher homotopy groups. The fiber $B^{k-1}\pi_k$ has a natural abelian group structure, and so can act vertically on the fibers of $X^k$. \\

The reason that the Whitehead tower is the natural setting is because here the solitonic charge configuration $Q$ has an elegant and convenient interpretation: \textit{it describes a $k$-connection for how to move along the fiber of the Whitehead stage $X^k$ when it is pulled back to a bundle over spacetime}. That is, if we pull back the Whitehead tower fibration to $M_d$ along the field map $\sigma : M_d \to X$, then the cohomology class $[Q]\in H^k(M_d,\pi_k)$ enforces the conditions for how any section should move along the fiber $B^{k-1}\pi_k$ as the fields $\sigma$ wind around the defect. Indeed, a higher $k$-connection $\gamma_k$ for a principal $k$-bundle with discrete abelian structure group $A$ over a space $M$ is defined by a cohomology class $\gamma_k \in H^k(M,A)$. We provide a more thorough argument for this perspective, as well as a brief review of the mathematical background, in the \hyperref[appa]{Appendix}.  \\

Now let us consider the electric symmetry, which originates from the $G$-action on $X$. For now we assume that $G$ is connected. Then for any $g\in G$ acting on $X$, there exists a lift to an action of $g$ on $X^k$ which commutes with the fibration map $ X^k \to X$. As a result, the vertical action of $B^{k-1} \pi_k$ on the fibers does not affect the $G$-action on $X^k$. Nevertheless, the lifted $G$ action can still \textit{mix} nontrivally with the vertical fiber action. In the figure below we illustrate an example of how this happens in the case of the compact boson with $X=\mathbb{S}^1$. The electric $U(1)$ symmetry around the circle lifts to an $\mathbb R$ translation on the universal cover $X^1=\mathbb R$. This implies that large $U(1)$ rotations of $X$ can mix with vertical translations of the fiber $\mathbb Z$ which labels the sheets of the cover, to form a nontrivial group extension $\mathbb Z \to \mathbb R\to U(1)$

\bigskip

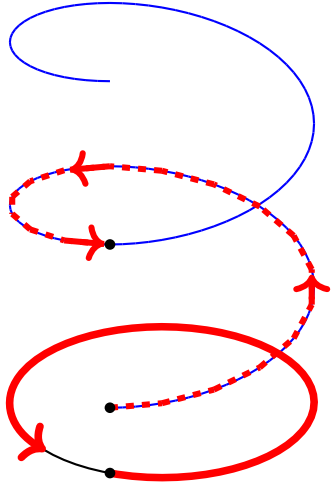
\begin{center} 
\begin{tikzpicture}
    \tdplotsetmaincoords{60}{110} 

    \begin{scope}[tdplot_main_coords]
        \def\radius{2}    
        \def\turns{2}     
        \def\height{5.0}  
        \def\gap{1.0}     
        \def\circleArc{330} 
        \def\spiralArc{360} 

        \draw[thick, black] (\radius,0,0) arc[start angle=0,end angle=360,radius=\radius];

        \foreach \t in {0,5,...,720} { 
            \pgfmathsetmacro\x{\radius*cos(\t)}
            \pgfmathsetmacro\y{\radius*sin(\t)}
            \pgfmathsetmacro\z{\gap + \height * \t / (360*\turns)} 

            \ifnum \t > 0
                \draw[thick,blue] (\xold,\yold,\zold) -- (\x,\y,\z);
            \fi

            \xdef\xold{\x}
            \xdef\yold{\y}
            \xdef\zold{\z}
        }

        \draw[line width=1mm,red,opacity=0.6,->] (\radius,0,0) arc[start angle=0,end angle=\circleArc,radius=\radius];

        \fill[black] (\radius, 0, 0) circle (2pt); 

        \foreach \t in {0,20,...,\spiralArc} { 
            \pgfmathsetmacro\x{\radius*cos(\t)}
            \pgfmathsetmacro\y{\radius*sin(\t)}
            \pgfmathsetmacro\z{\gap + \height * \t / (360*\turns)} 

            \ifnum \t > 0
                \draw[line width=0.8mm,red,dashed] (\xold,\yold,\zold) -- (\x,\y,\z);
            \fi

            \ifnum \t = 0 \relax
            \else
                \ifnum \t = 120 \relax
                    \draw[line width=0.8mm,red,->,shorten >=2pt,shorten <=2pt] (\xold,\yold,\zold) -- (\x,\y,\z);
                \fi
                \ifnum \t = 240 \relax
                    \draw[line width=0.8mm,red,->,shorten >=2pt,shorten <=2pt] (\xold,\yold,\zold) -- (\x,\y,\z);
                \fi
                \ifnum \t = 360 \relax
                    \draw[line width=0.8mm,red,->,shorten >=2pt,shorten <=2pt] (\xold,\yold,\zold) -- (\x,\y,\z);
                \fi
            \fi

            \xdef\xold{\x}
            \xdef\yold{\y}
            \xdef\zold{\z}
        }

        \fill[black] (\radius, 0, \gap) circle (2pt); 
        \fill[black] (\radius, 0, \gap + \height/2) circle (2pt); 

    \end{scope}
\end{tikzpicture}
\end{center} 

\noindent In general, this implies the possibility for the existence of a higher group extension \cite{Bunk:2020rju,Fiorenza:2013kqa}:
\begin{equation}\label{chextension}
    B^{k-1}\pi_k \to \Gamma^{(k)} \to G 
\end{equation}
Here we interpret $\Gamma^{(k)}$ as a $k$-group, since it is the extension of an ordinary 0-form symmetry by a $(k-1)$-form ``symmetry". This extension is classified by an extension class $[e] \in H^{k+1}(BG, \pi_k) $. The extension class is defined by taking the induced principal fibration on classifying spaces:
\begin{equation} \label{chextension}
    B^k\pi_k \to B\Gamma^{(k)} \to BG
\end{equation}
If we think of \hyperref[chextension]{(3.2)} as a $k$-bundle, then it is topologically classified by a classifying map $e : BG \to B^{k+1}\pi_k$, and this is what defines the cohomology class $[e]$. In general this extension can be nontrivial, and we will see that this is the root of the existence of the mixed electric-magnetic anomaly. \\

We can now reconcile two facts we have learned about the (pullback of the) Whitehead tower stage $X^k$: (i) there exists a possible nontrivial extension between $G$ and the group action along the fibers $B^{k-1}\pi_k$ and (ii) a solitonic charge configuration $Q$ may be interpreted as defining a $k$-connection, telling us how to move along these fibers. Considering these together, we find precisely the kind of extension between the electric symmetry and the homotopy charges that we were looking for. What the fibration \hyperref[chextension]{(3.2)} is telling us is that the act of coupling to a background electric gauge field $g : M_d\to BG$ and inserting a solitonic charge $Q : M_d \to B^k\pi_k$ are not independent operations: they are topologically connected. \textit{They should each only be considered as the components of a single connection $\gamma$ for the $k$-group} $\gamma : M_d \to B\Gamma^{(k)}$. In particular, large background gauge transformations of $G$ can induce a shift in the $k$-connection $Q$ which defines the solitonic charges, where this shift is determined by the extension class $e$. \\

Thus we have found the relationship between the electric symmetry $G$ and the magnetic charges $Q$, and seen that they are indeed related by an extension. It should be stressed that this is not an extension of symmetries, since $Q$ does not represent a connection for any symmetry of the theory. It is only an expression of the topological relationship between the magnetic charges and the background gauge field $g$ for $G$. Our next goal is to see how this extension leads to a mixed anomaly with the magnetic symmetry $M_k$. To do this we will need to introduce background gauge fields for this symmetry. \\

\subsection{The Anomaly from Charge Evaluation}
\label{sec33}

Now we are ready to introduce our magnetic symmetry $M_k = \hat \pi_k$ into the game. This is a $(d-k-1)$-form symmetry, so coupling to a background gauge field corresponds to introducing a $(d-k)$-form $\hat{\pi}_k$-connection, which is represented topologically by a map $ A_{d-k} : M_d \to B^{d-k} \hat\pi_k = B^{d-k}M_k$. We learned in the previous section that if we want to simultaneously couple to a background $G$ gauge field $g : M_d\to BG$ and insert a configuration of magnetic charges $Q : M_d\to B^{k}\pi_k$, then this required defining a connection on the bundle $B\Gamma^{(k)}$ defined by \hyperref[chextension]{(3.2)}. Thus, if we \textit{also} want to couple to a background $M_k$ gauge field $A_{d-k}$, this naively corresponds to defining a connection on the following bundle:
\begin{equation}
    B^k\pi_k \to B\Gamma^{(k)}\times B^{d-k}\hat \pi_k \to BG \times B^{d-k}\hat \pi_k
\end{equation}
This bundle is defined topologically by a classifying map $ (e,\text{id}) : BG\times B^{d-k}\hat \pi_k \to B^{k+1}\pi_k \times B^{d-k}\hat \pi_k$, where $\text{id}$ indicates that the right component is just the identity: 

\[\begin{tikzcd}
	{B^k\pi_k} & {B\Gamma^{(k)}\times B^{d-k}\hat\pi_k} \\
	& {BG\times B^{d-k}\hat\pi_k} & {B^{k+1}\pi_k\times B^{d-k}\hat \pi_k}
	\arrow[from=1-1, to=1-2]
	\arrow[from=1-2, to=2-2]
	\arrow["{(e,\text{id})}", from=2-2, to=2-3]
\end{tikzcd}\]

However we have not taken into account the fact that a configuration $Q$ represents the charges of the $M_k$-symmetry, and so can not be considered independent of the background field $A_{d-k}$. The connection $A_{d-k}$ can be thought of topologically as the insertion of a network of symmetry defect operators for $M_k$, and these defect operators \textit{act} on the charges $Q$ to produce a phase. They do this by linking the $A_{d-k}$ operators with the $Q$ operators and performing the evaluation map $\pi_k \times \hat\pi_k \to U(1)$. Indeed, the true connection for the magnetic symmetry, in the presence of these charges, should come from \textit{evaluating} the connection $A_{d-k}$ on the $Q$ defects. In terms of symmetry defect operators, this comes from linking the $\hat \pi_k$ defects with the $\pi_k$ defects, which should yield a phase $\exp(i\int_{M_d} A_{d-k}\cup Q)$.  \\

This evaluation procedure between connections $a_p : M_d \to B^p H$ and $b_{q} : M_d \to B^{q} \hat H $ for abelian group $H$ is implementable by a map on the classifying spaces:
\begin{equation}
    B^p H \times B^q \hat H \to B^p H \wedge B^q \hat H \to B^{p+q} U(1)
\end{equation}
where the first map is the quotient map of the smash product, and the second map is the composition of $K(H,p)\wedge K(\hat H, q) \cong K(H\otimes \hat H, p+q) \to B^{p+q}U(1)$ which is the map on classifying spaces induced by the evaluation map $H\times \hat H\to U(1)$. We denote this evaluation map as $f_{p,q} : B^p H \times B^q \hat H \to B^{p+q}U(1)$. \\

Applying this to the naive configuration of connections we described above, we see that the \textit{true} connections should be defined on a bundle which is obtained by extending the classifying map $ BG\times B^{d-k}\hat\pi_k \xrightarrow{(e,\text{id})} B^{k+1} \pi_k\times B^{d-k}\hat\pi_k$ by the map $f_{k+1,d-k}$. This defines the total classifying map $\alpha_{d+1} \equiv f_{k+1,d-k}\circ (e,\text{id}) : BG\times B^{d-k}M_k \to B^{d+1}U(1)$ coming from the composition of classifying maps:
\begin{equation}
    BG \times B^{d-k}\hat \pi_k \xrightarrow{(e,\text{id})} B^{k+1} \pi_k \times B^{d-k}\hat \pi_k \xrightarrow{f_{k+1, d-k}} B^{d+1}U(1)  
\end{equation}
The classifying map $\alpha_{d+1}$ defines a bundle $B^{d}U(1) \to B\mathbb G \to BG \times B^{d-k}\hat \pi_k$. Thus, after taking into account the action of $A_{d-k}$ on $Q$, we see that the true connections in our theory should be defined on $B\mathbb G$. We refer to this extension from the naive connections on $B\Gamma^{(k)}\times B^{d-k}\hat\pi_k $ to the true connections on $B\mathbb G$ as the \textbf{evaluation transformation}.   

\[\begin{tikzcd}
	{B^k\pi_k} & {B\Gamma^{(k)}\times B^{d-k}M_k} & \dashrightarrow & {B^dU(1)} & {B\mathbb G} & {} \\
	& {BG\times B^{d-k}M_k} & \dashrightarrow && {BG\times B^{d-k}M_k} & {}
	\arrow[from=1-1, to=1-2]
	\arrow[from=1-2, to=2-2]
	\arrow[from=1-4, to=1-5]
	\arrow[from=1-5, to=2-5]
\end{tikzcd}\]

\bigskip

Thus, we see that the true symmetry of our theory $\mathbb G$ is described by the higher group extension defined by the bundle on the right:
\begin{equation}\label{ext}
    U(1)^{(d-1)} \to \mathbb G \to G \times M_k^{(d-k-1)}
\end{equation}
This extension shows that there is a mixed anomaly between the electric symmetry $G$ and the magnetic symmetry $M_k$. The SPT phase for this anomaly is equivalent to the extension class $\alpha_{d+1} \in H^{d+1}(BG \times B^{d-k}M_k, U(1))$. This can be seen explicitly by extending $M_d$ to a $(d+1)$-dim bulk satisfying $M_d = \partial Y_{d+1}$ and coupling to background gauge fields $(g_1, A_{d-k}) : Y_{d+1} \to BG\times B^{d-k}M_k$. Then a gauge transformation of this connection can be canceled by inflow from a corresponding variation in the following bulk anomaly theory:
\begin{equation}\label{anom}
    \alpha_{d+1}[g_1,A_{d-k}] = \exp\Big( i \int_{Y_{d+1}} g_1^*(e) \cup A_{d-k} \Big)
\end{equation}

Notice that this is the same formula for the mixed anomaly found in \cite{Tachikawa:2017gyf} from gauging an abelian subgroup participating in a nontrivial extension with $G$ with extension class $e$. \\

An interpretation of a mixed 't Hooft anomaly in terms of an extension like \hyperref[ext]{(3.6)} by a $(d-1)$-form $U(1)$ ``symmetry" was discussed in \cite{Tachikawa:2017gyf}.  There, it was noted that every QFT possesses such a $U(1)^{(d-1)}$ symmetry which is just the ability to insert a point operator that contributes an overall phase to the partition function. A nontrivial extension $ 0\to U(1)^{(d-1)}\to \hat{\mathcal G} \to \mathcal G\to 0$ specifies the anomaly by its extension class in $H^{d+1}(BG, U(1))$, since a background gauge transformation of $G$ must then be accompanied by a phase determined by this extension class. \\

Thus, we see how an extension \hyperref[chextension]{(3.1)} of the electric symmetry $G$ by the homotopy charges $\pi_k^{(k-1)}$ leads to a mixed 't Hooft anomaly between $G$ and $M_k$ with SPT phase \hyperref[anom]{(3.7)}. \\

\subsection{A Twist from the Mapping Class Group}
\label{sec34}

So far we have made the simplifying assumption that the electric symmetry $G$ is connected. However in general this is not the case. The group of components of $G$ is known as the \textit{mapping class group} $\text{MCG}(X)\equiv \pi_0\text{Diff}(X)$ of the target space $X$. In order to understand the full extension of the electric symmetry and its effect on the anomaly, we need to include the contribution of $\text{MCG}(X)$. \\

In fact there is a well-known way in which the mapping class group interacts with the homotopy groups. In general there may be a group action of $\text{MCG}(X)$ on each $\pi_i(X)$. For example, in the case of target space $X=S^2$ then the full isometry group is $O(3)$, so the connected part is $SO(3)$ while the mapping class group is $\mathbb{Z}_2$. Indeed, the action of the mapping class group on $n\in \pi_2(S^2)$ takes $n \to -n$. We will typically denote this action $\rho_k : \text{MCG}(X) \to \text{Aut}(\pi_k(X))$.  \\

For the case of the lowest nontrivial homotopy group $\pi_k(X)$, there is a group which quantifies to what extent the action of $\text{MCG}(X)$ on $\pi_k$ is nontrivial. This is the \textit{Torelli group} $\text{Tor}(X)$, and it is defined as the kernel of the action:
\begin{equation}
    0 \to \text{Tor}(X) \to \text{MCG}(X) \to \text{Aut}(H_k(X)) \to 0
\end{equation}
where we have used the Hurewicz isomorphism $H_k(X)\cong \pi_k(X)$. As long as the Torelli group is a proper subgroup of $\text{MCG}(X)$ then the action is nontrivial. \\

The effect of this action on the extension \hyperref[chextension]{(3.1)} with the homotopy charges is to twist the extension $0 \to \pi_k^{(k-1)} \to \Gamma^{(k)} \to G\to 0$, so that the extension is now defined by the data $(e,\rho_k)$. Such a twisted extension also has a description in terms of classifying maps, except now the extension is defined by a classifying map $\tilde e : BG \to \bar K(\pi_k, k+1 ; \rho) $ into a classifying space with a local coefficient system which incorporates the action $\rho : \text{MCG}(X)\to \text{Aut}(\pi_k)$ \cite{CTGDC_2003__44_1_63_0}. We can then use the fact that the action $\rho$ induces an action $\tilde \rho : \text{MCG}(X) \to \text{Aut}(\hat \pi_k) $ such that $\tilde \rho(\hat a) = a\circ \rho$ for $a\in \pi_k$ and $\hat a \in \hat \pi_k$. Moreover this definition makes the action $\rho$ equivariant with respect to the evaluation pairing $\pi_k \times \hat \pi_k \to U(1)$, which in turn implies that there exists a twisted cup product $\cup_{\rho}$ between cohomology classes \cite{localcoef, Benini:2018reh}. We expect that there should exist an analogous pairing on the twisted classifying spaces $\bar K(A,n;\rho)$, although we do not seek a rigorous derivation here. \\

Thus we can follow through the same argument in the previous section, where now we use the classifying spaces $\bar K( \pi_k , k+1;)$ and $\bar K(\hat \pi_k, d-k; \rho)$ in place of $B^{k+1}\pi_k$ and $B^{d-k}\hat \pi_k$. In the end we expect a twisted extension:
\begin{equation}
    U(1)^{(d-1)} \to \mathbb G \to G \rtimes_{\rho} \hat\pi_k^{(d-k-1)} 
\end{equation}
so that the resulting symmetry is the semidirect product $G \rtimes_{\rho} \hat\pi_k $. We propose that the pairing on twisted classifying spaces should yield a mixed anomaly whose SPT phase is described by the extension class: 
\begin{equation}\label{twaf}
    \alpha_{d+1}[g_1,A_{d-k}] = \exp\Big( i \int_{Y_{d+1}} g_1^*(e) \cup_{\rho} A_{d-k} \Big)
\end{equation}

Thus we see that the effect of the $\text{MCG}(X)$ action is to twist the resulting symmetry into a semidirect product, as well as to twist the anomaly theory. \\

\newpage

\section{Examples Revisited}
\label{sec4}

Now that we are equipped with a new perspective on the mixed anomalies between our electric and magnetic symmetries, we can test drive them on our previous examples to see if they reproduce the same results. \\

\subsection{The Compact Boson}
\label{sec41}

The compact boson has target space $X= \mathbb S^1$. We found that the electric symmetry is the translation symmetry $G= U(1)$ while the $M_1=U(1)$ $(d-2)$-form magnetic symmetry came from $\pi_1(\mathbb S^1)=\mathbb Z$. In order to understand the extension between $G$ and the $\pi_1$-charges, we need to look at the first stage of the Whitehead tower, which is just the universal cover of $\mathbb S^1$. This is clearly just the line $\mathbb R$, and since this is contractible then the whole Whitehead tower is just the covering space $\mathbb Z \to \mathbb R \to \mathbb S^1$. \\

Therefore if we lift the electric $G=U(1)$ action on $\mathbb S^1$ to the universal cover, the action just becomes the translations of $\mathbb R$. This fits into the group extension with the fiber: $\mathbb Z \to \mathbb R \to U(1)$. The extension class is just the first Chern class $e = c_1 \in H^2(BU(1), \mathbb Z)$. The only element of the mapping class group $\text{MCG}(\mathbb S^1)$ is the antipode map, which acts trivially on $\pi_1(\mathbb S^1)$, so there is no twist. \\

We can apply our anomaly formula \hyperref[anom]{(3.7)} if we couple to background gauge fields $A_1$ for $G$ and  $B_{d-1}$ for $M_1$. We can express the pullback of the extension class as just the curvature of the gauge field: $A_1^*e = c_1(A_1)=dA/2\pi$. Then the anomaly becomes:
\begin{equation}
    \alpha_{SPT} = \exp\Big(i \int A_1^*e\cup B_{d-1}\Big)=\exp\Big(\frac{i}{2\pi} \int dA_1\wedge B_{d-1}\Big)
\end{equation}
which is precisely the anomaly theory \hyperref[CBanom]{(2.3)} that we found earlier. \\

\subsection{The Torus}
\label{sec42}

The case of the torus $X = \mathbb T^2$ carries through precisely the same, since this is just two copies of the compact boson. The full Whitehead tower is just the universal cover $\mathbb Z^2 \to \mathbb R^2 \to \mathbb T^2$. Electric symmetry transformations $G = U(1)^2$ get lifted to translations of $\mathbb R^2$, and thus fit in to an extension with the charges: $\mathbb Z^2 \to \mathbb R^2 \to U(1)^2$, classified by $(c_1,c_1)\in H^2(BU(1)^2, \mathbb Z^2)$. \\

However now there is the additional contribution from the mapping class group $\text{MCG}(\mathbb T^2)=SL(2,\mathbb Z)$, so that the full electric symmetry is $G = SL(2,\mathbb Z)\rtimes U(1)^2$. This acts on the homotopy charges with a nontrivial action $\rho : SL(2,\mathbb Z)\to \text{Aut}(\mathbb Z^2)$. \\

Therefore the true symmetry of the theory is twisted $U(1)^2\rtimes SL(2,\mathbb Z) \rtimes_{\rho} [U(1)^{(d-2)}]^2$, and the anomaly theory should be defined using the twisted anomaly \hyperref[twaf]{(3.10)}:
\begin{equation}
    \alpha_{SPT}=\exp\Big( i \int (c_1(A_1^1),c_1(A_1^2))\cup_{\rho} (B_{d-1}^1, B_{d-1}^2)\Big)=\exp\Big(\frac{i}{2\pi} \int \sum_i dA_1^i\wedge_{\rho} B_{d-1}^i\Big)
\end{equation}
This recovers the anomaly theory \hyperref[tanom]{(2.5)} that we proposed previously. \\

\subsection{Generalized Maxwell Theory}

The case of generalized Maxwell theory is just the higher-form generalization of the compact boson. It is not proper to think of this as a $\sigma$-model, but it is true that the topologically-distinct configurations of a $p$-form abelian gauge field are classified by homotopy maps into $B^pU(1)$, which is $p$-connected and so in some sense fits into a universal fibration $B^{p-1}U(1)\to E \to B^pU(1)$.  \\ 

It is interesting to compare our discussion here with the results of \cite{Freed:2006ya, Freed:2006yc}. In this work they found the Hilbert space associated to a spatial slice $Y$ could be expressed explicitly using differential cohomology, so that the gauge-invariant states associated to $p$-form field configurations could be represented by Cheeger-Simons differential characters $\check H^p(Y)$, and the Poincare-dual configurations with the group $\check H^{d-p}(Y)$. This allows for a natural definition of the electric and magnetic charge sectors, since a magnetic charge of a state $\ket \chi$ associated to a differential cocycle $\check \chi \in \check H^p(Y)$ can be defined by the cohomology class $m\in H^p(Y,\mathbb Z) \subset \check H^p(Y)$ required for the definition of $\check \chi$. This induces a natural grading on the Hilbert space into definite charge sectors $ m \in H^p(Y,\mathbb Z)$. On the other hand, the electric charge sectors can be labeled by $e \in H^{d-p}(Y,\mathbb Z)$. A state in such an electric charge sector is an eigenstate $\ket{\psi}$ under the $p$-form electric symmetry, which acts by shifting by a flat character $\check \phi \in H^{p-1}(Y,U(1))$, so that $\psi(\check \chi+\check \phi)=\exp\Big(2\pi i \int_Y e \cup \phi\Big)\psi(\check \phi)$ for $e \in H^{n-p}(Y,\mathbb Z)$. \\

Their key result is that there exists no simultaneous bigrading of the Hilbert space into electric and magnetic charge sectors, and instead the Hilbert space is the unique irrep of a Heisenberg group $\text{Heis}(\check H^p(Y)\times \check H^{d-p}(Y))$, which defines a projective representation between the electric and magnetic sectors with a cocycle associated to the link pairing on the cohomology classes. This obstruction to simultaneously decomposing into electric and magnetic charge sectors is indicative of the mixed anomaly between the electric and magnetic symmetries. \\

We can see how this perspective is related to the arguments we gave in \hyperref[sec3]{Section 3}. The fact that we can not avoid the projective representation between the electric and magnetic charge sectors can be interpreted as the fact that the electric symmetry does not leave the magnetic flux sectors untouched: performing a large electric transformation which translates by a topologically nontrivial flat character necessarily carries us to a different magnetic charge sector \cite{Freed:2006ya}. This is the same sort of extension between the electric symmetry $G$ and the magnetic charges $\pi_k(X)$ that we found in \hyperref[sec32]{Sec 3.2}. Moreover if we use our anomaly formula \hyperref[anom]{(3.7)} then we reproduce the usual mixed $p$-form electric-magnetic anomaly theory \hyperref[EManom]{(2.6)}. \\

\subsection{The $\mathbb{CP}^1$ Model (Round II)}

Let us revisit the 4d $\mathbb{CP}^1$ model. The electric symmetry is $G=O(3)$ while the lowest magnetic symmetry is the 1-form vortex symmetry $M_2= \hat \pi_2(\mathbb{CP}^1)=U(1)^{(1)}$. In order to understand the extension between $G$ and the magnetic charges we need to look at the second stage of the Whitehead tower of $\mathbb S^2$. This Whitehead fibration is in fact just the Hopf fibration $\mathbb S^1\to \mathbb S^3 \to \mathbb S^2$. \\

If we consider the connected piece of the electric symmetry $SO(3)$, then this lifts to an $SU(2)$ action on $\mathbb S^3$. If we consider a magnetic charge $A_n$ which is a vortex of charge $n$, then, as we go around the vortex, the field winds $n$ times around the $\mathbb S^1$ Hopf fiber of  $\mathbb S^3$. The translation along the Hopf fiber can mix with the lift of the $SO(3)$ action to yield a group extension:
\begin{equation}
    0 \to U(1) \to \Gamma \to SO(3) \to 0
\end{equation}
which is classified by the extension class $\beta w_2 \in H^2(BSO(3),U(1)) \cong H^3(BSO(3),\mathbb Z) $. If we couple to background gauge fields $A_1$ for $G$ and $B_2$ for $M_2$ then we can apply our anomaly formula \hyperref[anom]{(3.7)} to see that we expect a 5d anomaly theory:
\begin{equation}
    \alpha_{SPT} = \exp\Big(i\int \beta w_2(A_1)\cup B_2\Big)\cong \exp\Big(i\int w_2(A_1)\cup \beta B_2\Big)
\end{equation}
which is precisely the same anomaly \hyperref[cpanom]{(2.7)} we found from symmetry fractionalization of the vortices. \\

However now we must also take into account the fact that $SO(3)$ is not the full electric symmetry, since there is a $\mathbb{Z}_2$ mapping class group coming from the full $G = O(3)$. This transformation corresponds to the antipode map on the sphere, which has a nontrivial action $\rho \in \text{Aut}(\pi_2(\mathbb S^2))$ which takes $\rho(n)=-n$. This has the effect of twisting the total symmetry to be $G\rtimes_{\rho} M_2=O(3)\rtimes_{\rho} U(1)^{(1)}$, and it twists the above anomaly formula so that we must instead use the twisted cup product:
\begin{equation}
    \alpha_{SPT} = \exp\Big(i\int w_2(A_1)\cup_{\rho} \beta B_2\Big)
\end{equation}

\subsection{The $SU(N)$ Model}

We can reconsider the 4d $SU(N)$ model, which has electric symmetry $G = PSU(N)\rtimes \mathbb Z_2$ and magnetic 0-form symmetry $M_3=\hat\pi_3(SU(N))\cong U(1)$. To understand the extension between the electric symmetry and the magnetic charges we need to consider the 3rd stage of the Whitehead tower of $SU(N)$. For any compact simple and simply-connected Lie group $H$ this is an infinite-dimensional topological group called $\text{String}(H)$. For example, for the $\text{Spin}(n)$ Lie groups this is just the usual string group $\text{String}(n)$. Therefore we need to consider $\text{String}_{SU}(N)$ defined by the fibration $BU(1) \to \text{String}_{SU}(N)\to SU(N)$. \\

If we lift the $PSU(N)$ action on $SU(N)$ to $\text{String}_{SU}(N)$ then this can mix with the abelian group action of $BU(1)$ along the fibers, resulting in a higher group extension:
\begin{equation}
    0 \to BU(1) \to \Gamma \to PSU(N) \to 0
\end{equation}
which is classified by an extension class $c_2 \in H^4(BPSU(N),\mathbb Z)$. The $\mathbb Z_2$ action of the mapping class group of $SU(N)$ acts on the homotopy classes in $\pi_3(SU(N))$ trivially so there is no twist. Then if we couple to background gauge fields $A_1$ for $G$ and $B_1$ for $M_3$ then we can apply our anomaly formula \hyperref[anom]{(3.7)} so that we expect a 5d anomaly theory:
\begin{equation}
    \alpha_{SPT} = \exp\Big( i\int c_2(A)\cup B_1\Big)
\end{equation}
which matches the anomaly \hyperref[suanom]{(2.8)} that we previously found.

\subsection{The $SO(3)$ Model}

As another example, let us consider the case when the target space is $X= SO(3)$ in 4d for $N \geq 3$. The electric symmetries of this theory are given by the automorphisms $G = \text{Aut}(SO(3))$. The outer automorphisms of $SO(N)$ correspond to the symmetries of the Dynkin diagrams, so the automorphisms are:
\begin{equation}
    G = \begin{cases}
        SO(N) & N \text{ even} \\ 
        SO(N)\rtimes \mathbb{Z}_2 &  N\text{ odd}
        \end{cases}
\end{equation}
The lowest nontrivial homotopy group is $\pi_1(SO(3)) = \mathbb Z_2$. In 4d this indicates a 2-form magnetic symmetry $M_1$ whose charged solitons are surface defects. We can investigate the extension between the electric symmetry and the homotopy charges by lifting to the universal cover $\text{Spin}(N)$. The $G$-transformations of $SO(3)$ then lift to $SU(2)$ transformations. The extension between $G$ and $\pi_1$ is then just the extension $\mathbb Z_2\to SU(2)\to SO(3)$ defined by the extension class given by the Stiefel-Whitney class $e=w_2\in H^2(BSO(3),\mathbb Z_2)$. There is also a $\mathbb Z_2$ mapping class group coming from the outer automorphism of $SO(3)$ by reflection. Since this action swaps the orientation of paths, then it has a nontrivial action on $\pi_1=\mathbb Z_2$ which just exchanges the elements. Therefore there is a twist by this action $\rho : \text{Out}(SO(3))\to \pi_1(SO(3))$ in the above extension. By coupling to background gauge fields $g_1 $ for $G$ and $A_3$ for $M_1$ then the mixed anomaly between the electric and magnetic symmetries is described by the 5d anomaly theory:
\begin{equation}
    \alpha^{(1)}_{SPT}=\exp\Big(i\int w_2(g)\cup_{\rho} A_3\Big)
\end{equation}

\newpage

\section{Higher Homotopy and Non-Invertible Symmetries}
\label{sec5}

So far we have only considered the solitonic symmetries coming from the lowest nontrivial homotopy group $\pi_k(X)$, which we assume to be abelian. This gives us an invertible $(d-k-1)$-form symmetry described by the group $\hat \pi_k$. However a general target space $X$ will typically have multiple nontrivial homotopy groups. The natural question is then how do we generalize our results to describe mixed anomalies between the electric symmetry $G$ and the solitonic symmetries resulting from these higher homotopy groups? \\

The complication in addressing this question comes from the fact that the homotopy groups of a space will generally participate in extensions with each other, as described by the fibrations in the Postnikov tower \cite{Pace:2023kyi, Pace:2023mdo}. As we now review in \hyperref[sec51]{Sec 5.1}, the full charge algebra is described by a higher group $\mathbb G^{(d-1)}$ such that $B\mathbb G^{(d-1)} \cong X_{d-1}$, and the true solitonic symmetry is described by the higher fusion category $\text{Rep}^{d-1}(\mathbb G^{(d-1)})$. When these extensions are nontrivial, the resulting solitonic symmetries can be non-invertible. We are thus asking for a homotopy description of the mixed anomaly between the electric isometry symmetry and the (possibly non-invertible) higher solitonic symmetry. The full treatment of the symmetries therefore requires the technology of higher fusion categories. The appropriate mathematics allowing for such a description in the general case is still lacking, although there are descriptions for the finite case, such as in \cite{etingof2009fusioncategorieshomotopytheory}.\\

In this work, we do not attempt a fully general description of the anomalies for the non-invertible case. We instead aim for an understanding of the higher homotopy solitonic symmetries when they remain invertible, using extensions of the tools we have developed thus far. Nevertheless, we will see that the obstructions we run into allow us to identify the existence of mixed anomalies with the non-invertible solitonic symmetries, which makes the task of characterizing these anomalies an enticing goal for future work. \\

\subsection{Higher Magnetic Symmetry from Higher Homotopy}
\label{sec51}

If we want to analyze the solitonic symmetries coming from higher homotopy groups of the target space then there can be further subtleties. These occur in cases where the target space has nontrivial homotopy type and the different homotopy groups mix to form a non-split higher group. As pointed out in \cite{Chen:2022cyw}, the charged solitons for the lower homotopy groups can thus carry nontrivial winding number under the higher homotopy groups. In extreme cases, the higher-homotopy solitonic symmetry can become non-invertible, and the full symmetry must be described by a higher fusion category \cite{Chen:2023czk}. In \cite{Pace:2023mdo} it was shown that the full solitonic symmetry can be understood using the higher group built from the homotopy groups appearing in the Postnikov tower of the target space. We now review these perspectives. \\

If we are working in a spacetime $M_d$ of dimension $d$, then our solitonic symmetries come from the nontrivial homtopy classes of maps $[M_d,X]$, and therefore depend only on the homotopy $d$-type of $X$. That is, we only care about the homotopy groups $\pi_k(X)$ up to $k=d-1$. Traditionally, the different solitonic symmetries are just considered to be the invertible $(d-k-1)$-form symmetries $\hat \pi_k(X)$ for each $k$, whose charged objects are the associated species of soliton. This would suggest the total solitonic symmetry is a product of the Pontryagin dual groups $M = \prod_k M_k = \prod_k \hat \pi_k^{(d-k-1)}$. \\

However there is an immediate issue with this. In principle we can take any closed $k$-cycle $\Sigma_k \subset M_d$ to support a soliton and establish the winding boundary conditions around $\Sigma_k$ to take value in the homotopy classes $[\Sigma_k\times \mathbb{S}^{d-k-1}, X] $. However, in general there are \textit{more} homotopy classes in this set than just those corresponding to $[\mathbb S^{d-k-1}, X]\sim  \pi_{d-k-1}(X)$. This means that $\Sigma_k$ can support more charges than just those captured in $\pi_{d-k-1}(X)$. These extra homotopy classes descend from the higher homotopy groups, and thus in general we need to consider the whole homotopy $d$-type of $X$ in order to find all possible charges. \\

This can be studied systematically using the Postnikov tower of the target space $X$ \cite{Pace:2023mdo}: 
\[\begin{tikzcd}
	& X \\
	& \vdots \\
	{B^3\pi_3(X)} & {X_3} \\
	{B^2\pi_2(X)} & {X_2} \\
	{B\pi_1(X)} & {X_1}
	\arrow[from=1-2, to=2-2]
	\arrow[from=2-2, to=3-2]
	\arrow[from=3-1, to=3-2]
	\arrow[from=3-2, to=4-2]
	\arrow[from=4-1, to=4-2]
	\arrow[from=4-2, to=5-2]
	\arrow[from=5-1, to=5-2]
\end{tikzcd}\] \\
Each Postnikov stage $X_k$ captures the homotopy $k$-type of $X$, and fits into a fibration $B^k\pi_k(X) \to X_k \to X_{k-1}$. The stages are defined by their homotopy groups: $\pi_i(X_k) =\pi_i(X)$ for $i \leq k$ while $\pi_i(X_k)=0$ for $i > k$. Each fibration is classified by a cohomology class $[\alpha_k] \in H^{k+1}(X_{k-1},\pi_k)$ corresponding to a classifying map $\alpha_k : X_{k-1}\to B^{k+1}\pi_k(X)$. This sequence of fibrations can be thought of as defining a sequence of group extensions with $\pi_k$ treated as a $(k-1)$-form ``symmetry". This means that the full algebra of solitonic charges up to $k$-dim is described by this higher group $\mathbb G^{(k)}$, and that the $k^{th}$ Postnikov stage is homotopy equivalent to the classifying space of this higher group $X_k = B\mathbb G^{(k)}$ \cite{Pace:2023mdo}. \\

Thus the full algebra of homotopy charges is not just the product $\prod_k\pi_k(X)$, but instead is described by the (possibly non-split) $(d-1)$-group $\mathbb G^{(d-1)} $. If all the homotopy groups are finite, this means that the true magnetic symmetry of the theory is given by the higher fusion category of $(d-1)$-representations of this higher group: $M = \text{Rep}^{d-1}(\mathbb G^{(d-1)})$. If the Postnikov classes $\alpha_k$ describing the extensions in $\mathbb G^{(d-1)}$ are nontrivial then this symmetry can be non-invertible. If any of the homotopy groups which participate in a nontrivial extension are infinite then the resulting symmetry will be a continuous non-invertible symmetry, for which the correct mathematical description is not known. \\

 It was proposed in \cite{Chen:2023czk}  that in the case of all finite homotopy groups, the full solitonic symmetry of $X$ can be understood by constructing a ``solitonic cohomology" theory, where the fusion $n$-category $\text{Rep}^n(X) \equiv \text{Fun}(X, \Sigma^{n-1}\text{Vect})$ is defined as a contravariant functor between $(\infty,n)$-categories. While we do not use this formalism here, we expect that it may be helpful in better understanding the anomalies of the solitonic symmetry in the non-invertible case. 

\subsubsection{The $\mathbb{CP}^1$ Model (Round III)}

The 4d $\mathbb{CP}^1$ model was the first example where non-invertible solitonic symmetry was understood \cite{Chen:2022cyw}, as well as being later studied using the Postnikov tower perspective in \cite{Pace:2023mdo}. We quickly review their findings as an instructive example. \\

As we found before, the theory has an invertible 1-form $U(1)$ vortex symmetry coming from the $\pi_2$. The topological symmetry defect operators which generate this symmetry can be written explicitly in terms of the solitonic current as $V_{\alpha}(\Sigma_2)=\exp(i\alpha \int_{\Sigma_2}dA/2\pi)$. We also anticipated that there should be a 0-form $U(1)$ symmetry coming from the $\pi_3=\mathbb Z$ whose charged solitons are Hopfions. We expect traditionally that the conserved current should be the Hopf invariant $\frac{1}{4\pi} AdA$ which measures the $\pi_3$-winding, and therefore the symmetry defect operators should be $U_{\alpha}(\Sigma_3)=\exp(i\alpha\int_{\Sigma_3}AdA/4\pi)$. However there is an immediate problem since this current is not gauge invariant. \\

It was pointed out in \cite{Chen:2022cyw} that this issue can be understood as equivalent to the fact that the $\pi_2$ vortices can carry $\pi_3$ Hopfion number. Indeed, we can define a vortex supported on a circle $\mathbb S^1$ by enforcing the winding boundary conditions on the boundary $\mathbb S^2\times \mathbb S^1 $ surrounding the vortex carry nontrivial homotopy class in $[\mathbb S^2\times \mathbb S^1, \mathbb{CP}^1]$. The usual vortices of charge $n$ are just taken to have winding $ n \in [\mathbb S^2, \mathbb {CP}^1]\cong \pi_2(\mathbb{CP}^1)\cong \mathbb Z$. However there are actually more possible charges since the full set of homotopy classes is \cite{Chen:2023czk}:
\begin{equation}
    [\mathbb S^2\times \mathbb S^1,\mathbb{CP}^1] \cong \{ (n,m)  : n\in \mathbb Z, m\in \mathbb Z_{|2n|}\}
\end{equation}
Therefore a vortex with $\pi_2$-charge $n$ actually has $|2n|$ more configurations coming from the $\pi_3$ symmetry charges. We can therefore ask what is the true $\pi_3$ symmetry which is capable of detecting the charge of both the Hopfions and the vortices? We can notice that the lack of gauge invariance of the $\pi_3$ current is reminiscent of the same issue in the would-be current of the broken axial symmetry in the ABJ anomaly \cite{Cordova:2022ieu, Choi:2022jqy}. In \cite{Chen:2022cyw} the authors propose that the true symmetry defect operators should be given by the 3d minimal abelian TQFTs $U_{p/N}(\Sigma_3)=\mathcal A^{N,p}[\Sigma_3]$ which support a fractional quantum Hall state \cite{Hsin:2018vcg}. These operators generate a non-invertible $\mathbb{Q/Z}$ symmetry. They demonstrate that these operators are capable of detecting the correct charge both by linking Hopfions and surrounding vortices. Thus we see that the presence of a vortex of charge $n$ modulates the invertible part of the Hopfion symmetry down to $\mathbb Z_{|2n|}$, and only a global $\mathbb Z_2$ is truly invertible. \\

It is interesting to point out that this might be considered the origin of the constraint that the 3d $\mathbb{CP}^1$ model is only allowed a $\theta$-term with $\mathbb Z_2$ coefficients \cite{Freed:2017rlk}. The $\theta$-term in 3d $\mathbb{CP}^1$ model is given by the 3-form Hopf current integrated over spacetime, so it weights the sum over topological sectors in the partition function by this Hopf number multiplied by $\theta$. Gauge invariance requires $\theta \in 2\pi\mathbb Z$, but we now see from the fact that the $\pi_2$ and $\pi_3$ mix to form a higher group that this Hopf number is dependent on the total $\pi_2$ charge present. Since only a total $\mathbb Z_2$ symmetry is invertible, then the only Hopf charge which is well defined for \textit{any} $\pi_2$ charge operator content is the Hopf number mod 2. This restricts the value of $\theta$ to only 0 or $\pi$.  \\

In \cite{Pace:2023mdo} the non-invertibility of the $\pi_3$-symmetry was shown to be seen immediately from the higher group of homotopy charges in the Postnikov tower of $\mathbb{CP}^1$. The lowest Postnikov fibration is given by $B^3\pi_3 \to \mathbb S^2_3\to \mathbb S^2_2$. Since $\mathbb S^2$ is simply connected then $\mathbb S^2_2\cong B^2\pi_2\cong K(\mathbb Z,2)$. The Postnikov invariant which classifies this fibration is given by $x_2\cup x_2\in H^4(K(\mathbb Z,2),\mathbb Z)$ where we have used $H^*(K(\mathbb Z,2),\mathbb Z)=\mathbb Z[x_2]$. This is the Postnikov class for the 3-group of homotopy charges $ \pi_3^{(2)} \to \mathbb G^{(3)} \to \pi_2^{(1)} $, and since it corresponds to an unstable cohomology operation then the dual symmetry $\text{Rep}^3(\mathbb G^{(3)})$ is non-invertible. 

\subsubsection{The $SO(3)$ Model}

Let us consider the 4d $\sigma$-model with target space $X=SO(3)$. The discussion can be readily generalized to $SO(N)$ but we take $SO(3)$ for simplicity. There is a 2-form magnetic symmetry $M_1$ resulting from $\pi_1(SO(3))\cong \mathbb Z_2$, whose charged solitons are surface defects. However in 4d there is also a 0-form symmetry $M_3=U(1)$ resulting from $\pi_3(SO(3))= \mathbb Z$, whose charged solitons are skyrmions. \\ 

To understand the relationship between these magnetic symmetries, we must look at the charge algebra in the Postnikov tower of $SO(3)$. Since $\pi_1(SO(3))=\mathbb Z_2$ then the lowest Postnikov stage is $SO(3)_1=B\mathbb Z_2$. The next nontrivial Postnikov stage is the third $SO(3)_3$ which is defined by the fibration $B^3\mathbb Z\to SO(3)_3\to B\mathbb Z_2$, and is defined by a Postnikov class $k_4 \in H^4(B\mathbb Z_2,\mathbb Z)
$ represented by a classifying map $ k_4 : B\mathbb Z_2\to B^4\mathbb Z$. This Postnikov class can be determined by the same method used for $\mathbb{CP}^1$ in \cite{Pace:2023mdo}, by taking the extension $\mathbb Z_2\xrightarrow{f} SU(2)\xrightarrow{\pi}SO(3)$ and projecting down onto the third Postnikov stages to yield a commuting diagram: 
\[\begin{tikzcd}
	{SU(2)} & {SO(3)} & {\mathbb{RP}^4} & {\mathbb S^4} \\
	{B^3\mathbb Z} & {SO(3)_3} & {B\mathbb Z_2} & {B^4\mathbb Z}
	\arrow["\pi", from=1-1, to=1-2]
	\arrow["{p_3}", from=1-1, to=2-1]
	\arrow["\subset", from=1-2, to=1-3]
	\arrow["{p_3}", from=1-2, to=2-2]
	\arrow["q", from=1-3, to=1-4]
	\arrow["{p_3}", from=1-3, to=2-3]
	\arrow["{p_4}", from=1-4, to=2-4]
	\arrow[from=2-1, to=2-2]
	\arrow[from=2-2, to=2-3]
	\arrow["k_4"', from=2-3, to=2-4]
\end{tikzcd}\]
We then use the fact that $H^*(B\mathbb Z_2,\mathbb Z)=\mathbb Z[x_2]$ with $p_3 \cong x_2$, along with the fact that the induced sequence in homotopy from the above extension does not split, so that $p_4\circ q$ is nontrivial, to conclude that $k_4 \cong x_2\cup x_2$. \\

Thus the charge algebra forms a 3-group $\pi_3^{(2)}\to \mathbb G^{(3)}\to \pi_1 $ with nontrivial Postnikov class $k_4$. Since this class represents an unstable cohomology operation, then the resulting solitonic symmetry $\text{Rep}^3(\mathbb G^{(3)})$ represents a non-invertible symmetry. \\

\subsubsection{$SO(3)$ Gauge Theory}

Let us consider $SO(3)$ gauge theory in 5d. The $SO(3)$ gauge fields are connections on a $SO(3)$ principal bundle, which is topologically classified by a classifying map into $BSO(3)$. The homotopy classes of $BSO(3)$ can therefore lead to magnetic symmetries. There exists a 2-form magnetic symmetry $M_2 = \mathbb Z_2$ resulting from $\pi_2(BSO(3)) = \mathbb Z_2$. There is also a 0-form magnetic symmetry $M_4= U(1)$ resulting from $\pi_4(BSO(3))\cong \mathbb Z$ for $N\neq 4$, whose charged objects are instantons.\\

Once again, the relationship between the magnetic symmetries can be seen from the lowest nontrivial fibration of the Postnikov tower, given by $B^4\pi_4 \to BSO(3)_4\to B^2\pi_2$. In particular, this is classified by a Postnikov class $k_5 \in H^5(B^2\mathbb Z_2,\mathbb Z)$. This class has been computed \cite{antieau2013classificationoriented3planebundles}, and it is the composition $k_5 = \delta \frak P$ of the Pontryagin square and the coboundary map:
\begin{equation}
    H^2(-,\mathbb Z_2)\xrightarrow{\frak P} H^4(-,\mathbb Z_4)\xrightarrow{\delta} H^5(-,\mathbb Z)
\end{equation}
The fact that this is not a stable cohomology operation implies that the instanton symmetry $M_4$ is non-invertible. As with the $\mathbb{CP}^1$ model we can see this extension manifest in the $\theta$-term in one-lower dimension. 4d $SO(3)$ gauge theory has a possible $\theta$-term which weights the sum over topological sectors in the partition function by instanton number. For $SO(3)$ there is an additional discrete choice including $\frac{\pi}2 \frak {P}(w_2)$ to form a generalized $\theta$-term \cite{Aharony:2013hda}. This choice amounts to a distinction between the theories $SO(3)_{\pm}$ which have distinct 't Hooft line operators.

\subsection{Higher Charge Extension and Evaluation}
\label{sec52}

Once again we begin trying to understand the anomaly by studying the topological relationship between the electric symmetry $G$ and the full algebra of homotopy charges. The generalization of the argument in \hyperref[sec32]{Sec 3.2} is straightforward: instead of just lifting the action of $G$ on $X$ to the lowest nontrivial $X^k$, we lift to each stage of the Whitehead tower. 
\[\begin{tikzcd}
	& \vdots \\
	{B^k\pi_{k+1}} & {X^{k+1}} \\
	{B^{k-1}\pi_{k}} & {X^k} \\
	& X
	\arrow[from=1-2, to=2-2]
	\arrow[from=2-1, to=2-2]
	\arrow[from=2-2, to=3-2]
	\arrow[from=3-1, to=3-2]
	\arrow[from=3-2, to=4-2]
\end{tikzcd}\]
Performing successive lifts defines a higher group extension of the electric symmetry $G$ by the higher homotopy charges $B^{k-1}\pi_k$. These successive extensions define a higher group $\Gamma^{(n)}$, so that each stage fits into an extension:
\[\begin{tikzcd}
	& \vdots \\
	{\pi_{k+1}^{(k)}} & {\Gamma^{(k+1)}} \\
	{\pi_k^{(k-1)}} & {\Gamma^{(k)}} \\
	& G
	\arrow[from=1-2, to=2-2]
	\arrow[from=2-1, to=2-2]
	\arrow[from=2-2, to=3-2]
	\arrow[from=3-1, to=3-2]
	\arrow[from=3-2, to=4-2]
\end{tikzcd}\]

We can define this sequence of extensions using the induced fibration on classifying spaces. Then each successive fibration is defined by an extension class $[e_n]\in H^{n+1}(B\Gamma^{(n-1)},\pi_n)$, determined by a classifying map $e_n : B\Gamma^{(n-1)} \to B^{n+1}\pi_n$. 

\[\begin{tikzcd}
	& \vdots \\
	{B^{k+1}\pi_{k+1}} & {B\Gamma^{(k+1)}} \\
	{B^k\pi_k} & {B\Gamma^{(k)}} & {B^{k+2}\pi_{k+1}} \\
	& BG & {B^{k+1}\pi_k}
	\arrow[from=1-2, to=2-2]
	\arrow[from=2-1, to=2-2]
	\arrow[from=2-2, to=3-2]
	\arrow[from=3-1, to=3-2]
	\arrow["{e_{k+1}}", from=3-2, to=3-3]
	\arrow[from=3-2, to=4-2]
	\arrow["{e_k}", from=4-2, to=4-3]
\end{tikzcd}\]

Thus the extension between the electric symmetry $G$ and the homotopy charges $\pi_n$ up to dimension $d-1$ is encoded in a $d$-group $\Gamma^{(d)}$. We would like to generalize the argument in \hyperref[sec33]{Sec 3.3}, which allowed us to include the magnetic symmetry $M_k=\hat \pi_n^{(d-k-1)}$ in the lowest fibration by coupling to a background connection $A_{d-k}: M_d\to B^{d-k}\hat\pi_k$. The true symmetry of the theory was then obtained by extending the classifying map defining the charge extension class $e_k$ using the evaluation pairing:
\begin{equation}
    B\Gamma^{(n-1)} \xrightarrow{(e_n,\text{id})} B^{n+1}\pi_n \times B^{d-n}\hat\pi_n \xrightarrow{f_{k+1,d-k}} B^{d+1} U(1)
\end{equation}
This evaluation transformation goes through precisely the same, if we perform it on the lowest fibration of the Whitehead tower. We end up with a description of the electric and magnetic symmetries as being extended by $U(1)^{(d-1)}$:

\[\begin{tikzcd}
	{B^k\pi_k} & {B\Gamma^{(k)}} && {} & {} & {B^dU(1)} & {B\mathbb G_k} \\
	& BG & {B^{k+1}\pi_k} & {} & {} && {BG\times B^{d-k}\hat\pi_k} & {B^{d+1}U(1)}
	\arrow[from=1-1, to=1-2]
	\arrow[from=1-2, to=2-2]
	\arrow[dashed, from=1-4, to=1-5]
	\arrow[from=1-6, to=1-7]
	\arrow[from=1-7, to=2-7]
	\arrow[from=2-2, to=2-3]
	\arrow[dashed, from=2-4, to=2-5]
	\arrow[from=2-7, to=2-8]
\end{tikzcd}\]

In order to understand the mixed anomalies with the higher homotopy charges, the most obvious thing to try would be to perform this evaluation procedure for every fibration in the tower which defines $\Gamma^{(d)}$. However there is an obvious obstruction to doing this: once we have performed this evaluation transformation at one level of the tower, it would seem that all higher levels are no longer defined. Indeed, we can see that the second fibration in the charge extension tower is defined by the classifying map $e_{k+1} : B\Gamma^{(k)}\to B^{k+2}\pi_{k+1}$. However once we have performed the evaluation transformation on the lowest fibration, we have transformed $B{\Gamma}^{(k)}\dashrightarrow B\mathbb G_k $, and there is no obvious way in which there exists an inherited map $B\mathbb G_k \to B^{k+2}\pi_{k+1}$ which ``preserves" $e_{k+1}$. \\

However we can circumvent this by realizing that we don't actually need to define a map $B\mathbb G_k \to B^{k+2}\pi_{k+1}$, since we eventually wanted to extend such a map anyways by the evaluation transformation. What we \textit{really} want is a map $E_{k+1}:B\mathbb G_k \to B^{k+2}\pi_{k+1}\wedge B^{d-k-1}\hat\pi_{k+1} \to B^{d+1} U(1)$ which suitably generalizes the map $e_{k+1}$. \\

Let us investigate the existence of such a map. This first requires a better understanding of the map $e_{k+1} : B\Gamma^{(k)}\to B^{k+2}\pi_{k+1}$, which defines a class $[e_{k+1}]\in H^{k+2}(B\Gamma^{(k)},\pi_{k+1})$. We can use the fibration $B^k\pi_k \to B\Gamma^{(k)} \to BG$ to decompose this cohomology class using the Serre spectral sequence:
\begin{equation}\label{5.2}
    E_2^{p,q} = H^p(BG , H^q(B^k \pi_k, \pi_{k+1})) \Rightarrow H^{p+q}(B\Gamma^{(k)}, \pi_{k+1})
\end{equation}
Thus, after taking the differentials into account, the class $e_{k+1}$ can be viewed as originating from a combination of elements in the groups $H^p(BG , H^q(B^k \pi_k, \pi_{k+1})$ such that $p+q=k+2$. \\

Now let us consider the hypothetical map $E_{k+1}$. We begin with the simple case where $\pi_k$ and $\pi_{k+1}$ are finite abelian groups, since in this case we have $B^{k+2}\pi_{k+1}\wedge B^{d-k-1}\hat\pi_{k+1}\cong B^{d+1}\pi_{k+1}$. Thus we are looking for a class $[E_{k+1}]\in H^{d+1}(B\mathbb G_k, \pi_{k+1})$ which is obtained from $e_{k+1}$ by the evaluation pairing. Once again we can use the defining fibration $B^{d}\pi_{k} \to B\mathbb G_k \to BG \times B^{d-k}\hat \pi_k$ and apply the spectral sequence:
\begin{equation}\label{5.3}
    E_2^{p,q} = H^p(BG\times B^{d-k}\hat\pi_k , H^q(B^d \pi_k, \pi_{k+1})) \Rightarrow H^{p+q}(B\mathbb G_k, \pi_{k+1})
\end{equation}
The key difference between \hyperref[5.2]{(5.2)} and \hyperref[5.3]{(5.3)} is that the coefficient group has changed from $H^q(B^k\pi_k,\pi_{k+1})$ to $H^q(B^d\pi_k,\pi_{k+1})$. It is possible to recover an element from $H^q(B^k\pi_k,\pi_{k+1})$ in $H^l(B^d\pi_k,\pi_{k+1})$ if this element corresponds to a stable cohomology operation, so that it can be mapped from an element of $H^l(B^d\pi_k,\pi_{k+1})$ by a series of cohomology suspensions. Otherwise however there is no general statement we can make to relate these.  \\

We consider this failure to reconstruct the higher stages of the extension as a signal that the symmetry becomes non-invertible. This is consistent with the claim in \cite{Pace:2023mdo} that a Postnikov class in the homotopy charge algebra which corresponds to an unstable cohomology operation indicates a non-invertible symmetry in the dual representation fusion category. In this case, then a homotopy description using the full fusion category is needed. A potential candidate might be the spectrum discussed in \cite{Chen:2023czk}. \\

However, in the event that we are working with a class $e_{k+1}$ which \textit{does} correspond to a stable cohomology operation, then this means that we can in fact consider a corresponding map $E_{k+1}:B\mathbb G_k \to B^{k+2}\pi_{k+1}\wedge B^{d-k-1}\hat\pi_{k+1} \to B^{d+1} U(1)$ which descends to $e_{k+1}$ under repeated cohomology suspension. This means that we can use this map to construct at least part of the next stage of the extension tower by the evaluation procedure. \\

Under repeated application of this evaluation procedure at each stage, we expect that all invertible parts of the solitonic symmetry should be recovered, allowing us to study their mixed anomalies. 

\subsection{Example: The $\mathbb{CP}^1$-Model (Round IV)}

As an example, let us once again consider our old friend the 4d $\mathbb{CP}^1$ model. We found earlier that the electric symmetry is $G= O(3)$ while there is a 1-form magnetic symmetry $M_2= \hat\pi_2=U(1)$. There is also a 0-form magnetic symmetry $M_3$ coming from $\pi_3=\mathbb Z$, however this symmetry is non-invertible due to the mixing between $\pi_2$ and $\pi_3$ in the Postnikov tower. The full charge algebra is then a 3-group $\mathbb G^{(3)}$ which sits in an extension $\pi_3^{(2)} \to \mathbb G^{(3)} \to \pi_2^{(1)}$ defined by Postnikov class $k_4 = x_2\cup x_2 \in H^4(B^2\mathbb Z,\mathbb Z)$.  \\

We study the mixed anomalies between the electric and magnetic symmetries by lifting to higher stages of the Whitehead tower. This was worked out for the lowest magnetic symmetry $M_2$ in \hyperref[sec4]{Section 4}, which involved lifting the $G=O(3)$ action on $\mathbb S^2$ to the 2-connected cover described by the Hopf fibration $B\mathbb Z\to \mathbb S^3\to \mathbb S^2$. This defined the 2-group $B\mathbb Z \to \Gamma^{(2)} \to G $. Now if we want to include the $M_3$ symmetry we need to lift to the third stage of the Whitehead tower, which is given by the string group $\text{String}(3)$, which fits into the fibration $B^2\mathbb Z\to \text{String}(3) \to SU(2)$. Therefore the total extension of $G$ by the homotopy charge algebra is found by lifting the $G=O(3)$ action to $\text{String}(3)$ and studying its extension with the fiber $B^2\mathbb Z$. This forms a 3-group $\Gamma^{(3)}$: 
\[\begin{tikzcd}
	{B^2\mathbb Z} & {\Gamma^{(3)}} \\
	{B\mathbb Z} & {\Gamma^{(2)}} \\
	& G
	\arrow[from=1-1, to=1-2]
	\arrow[from=1-2, to=2-2]
	\arrow[from=2-1, to=2-2]
	\arrow[from=2-2, to=3-2]
\end{tikzcd}\]
We found the mixed anomaly between $G$ and $M_2$ by performing the evaluation transformation on the lowest fibration: 
\[\begin{tikzcd}
	{B^2\mathbb Z} & {B\Gamma^{(2)}\times B^{2}M_2} & \dashrightarrow & {B^4U(1)} & {B\mathbb G^{(2)}} & {} \\
	& {BSO(3)\times B^{2}M_2} & \dashrightarrow && {BSO(3)\times B^{2}M_2} & {}
	\arrow[from=1-1, to=1-2]
	\arrow[from=1-2, to=2-2]
	\arrow[from=1-4, to=1-5]
	\arrow[from=1-5, to=2-5]
\end{tikzcd}\]
with the resulting bundle being classified by $\exp(\beta w_2\cup \text{id})  \in H^5(BSO(3)\times B^2M_2,U(1))$. However now we need to investigate the possibility of recovering the next stage $\Gamma^{(3)}$ of the extension. This was defined by a classifying map $e_4 : B\Gamma^{(2)}\to B^4\mathbb Z$, but we want to perform the evaluation transformation, so what we are really looking for is a map $E_4: B\mathbb G^{(2)}\times B M_3 \to B^5 U(1)$ which is obtained by evaluating $e_4$. \\

This first requires better understanding the class $e_4 \in H^4(B\Gamma^{(2)},\mathbb Z)$. We can study it using the Serre spectral sequence for the fibration $B^2\mathbb Z \to B\Gamma^{(2)}\to BSO(3)$:
\begin{equation}
    E_2^{p,q} = H^p(BSO(3), H^q(B^2\mathbb Z,\mathbb Z))\Rightarrow H^{p+q}(B\Gamma^{(2)},\mathbb Z)
\end{equation}
The possible contributions to $H^4(B\Gamma^{(2)},\mathbb Z)$ therefore come from $H^4(BSO(3),\mathbb Z)$ and $H^4(B^2\mathbb Z,\mathbb Z)$. The obstruction to lifting from $\text{Spin}(3)$ to $\text{String}(3)$ is $\frac12 p_1$ where $p_1\in H^4(BSO(3),\mathbb Z)$ is the first Pontryagin class. Moreover, we saw in \hyperref[sec51]{Sec 5.1} that $x_2\cup x_2 \in H^4(B^2\mathbb Z,\mathbb Z)$ was the Postnikov class which defines the higher group between the $\pi_2$ and $\pi_3$ charges. One can easily check that there are no differentials in the spectral sequence afflicting these groups, so therefore we expect that $e_4 \sim (\frac 12 p_1, [x_2]^2)\in H^4(B\Gamma^{(2)},\mathbb Z)$. \\

Now what are the possible classes that might define $E_4 \in H^5(B\mathbb G^{(2)}\times BM_3, U(1))$? We can once again decompose this using the Serre spectral sequence: 
\begin{equation}
    E_2^{p,q} = H^p(BSO(3)\times B^2M_2, H^q(B^4U(1),\mathbb Z))\Rightarrow H^{p+q}(B\mathbb G^{(2)},\mathbb Z)
\end{equation}
To find a class which is obtained by linking with the background gauge field $A : M_d\to BM_3$ then we need $p+q=4$. The only possible contribution comes from $\frac 12 p_1$ in $H^4(BSO(3)\times B^2M_2,\mathbb Z)$, which allows for a class $E_4 = \exp( \frac i2 p_1\cup \text{id}) \in H^5(B\mathbb G^{(2)}\times BM_3,U(1))$. We can express this in terms of background gauge fields $g : M_d\to BSO(3)$ for $G$ and $B : M_d\to BM_3=BU(1) $ for $M_3$ as the 5d anomaly theory:
\begin{equation}
    \alpha_{SPT}=E_4= \exp\Big(\frac{i}{2}\int_Y p_1(g)\wedge B\Big)
\end{equation}
This clearly captures the invertible part of the anomaly, coming from lifting the electric $SO(3)$ action to $\text{String}(3)$. However we can not consider this as a true full description of the anomaly theory, since the map $E_4$ fails to incorporate the class $x_2\cup x_2\in H^4(B^2\mathbb Z,\mathbb Z)$ which encodes the mixing between the $\pi_2$ and $\pi_3$ when we lift to $\text{String}(3)$. This failure originates from the fact that this class does not correspond to a stable cohomology operation, and thus does not lie in the image of the cohomology suspension from higher classifying spaces for $\pi_2$, which is the signature of a non-invertible solitonic symmetry. In this sense, $E_4$ is not the correct map to use to characterize the anomaly, since it does not truly capture all of $e_4$, and thus the above SPT phase can not be considered complete. Nevertheless, the fact that the extension class $e_4$ contains a nontrivial contribution from this non-invertible symmetry indicates that there does exist a mixed anomaly between this symmetry and the electric symmetry $G$. \\

The full characterization of the anomaly would seem to require a homotopy classification of the higher fusion category $\text{Rep}^3(\mathbb G^{(3)})$. In \cite{Chen:2022cyw} it was shown that this symmetry includes the non-invertible  $\mathbb{Q/ Z}$  symmetry which appears in many other 4d gauge theories \cite{Cordova:2022ieu, Choi:2022jqy}. The mathematical tools for such a description are still unknown. \\

\section{Discussion}
\label{sec6}

In this work we have successfully diagnosed the existence of a 't Hooft anomaly in bosonic $\sigma$-models between electric and magnetic symmetries as arising from a nontrivial extension in the Whitehead tower of the target space. When both symmetries are preserved, then the calculation of these anomalies reduces to a topological analysis of this Whitehead tower and the successive lifts of the electric symmetry. We also found that a nontrivial mapping class group for the target space can twist both the magnetic symmetry and the resulting anomaly. We showed that the analysis can be extended to magnetic symmetries for higher homotopy groups as long as they remain invertible, and otherwise is capable of detecting the existence of mixed anomalies when they are non-invertible. While this analysis gives powerful tools for determining anomalies, many interesting theories possess additional structure that may contribute to this analysis. These augmentations allow for many exciting future directions to expand this work. \\

The most immediate issue, which we discussed in \hyperref[sec5]{Section 5}, is to discover the correct framework to characterize the anomalies for magnetic symmetries corresponding to higher homotopy groups when the full symmetry becomes non-invertible. The task of understanding anomalies of non-invertible symmetries is an active area of research \cite{Kaidi:2023maf, Apruzzi:2023uma, Bartsch:2023wvv, Choi:2023xjw, Zhang:2023wlu, Cordova:2023bja, Anber:2023pny, Antinucci:2023ezl, Seifnashri:2024dsd}. The most natural generalization of the approach discussed here would appear to ask for a homotopy description of non-invertible symmetries. In \cite{Chen:2023czk} they propose a cohomology theory with TQFT coefficients for studying solitonic symmetries, however they note there are challenges when homotopy groups are infinite. \\

A popular proposal for a unified description of all symmetries in a quantum field theory is to use the bulk $(d+1)$-dim SymTFT. Indeed, it was shown in \cite{Pace:2023mdo} that the magnetic symmetries could conveniently be understood from this perspective. It would be interesting to see how the electric symmetries could also be incorporated into the SymTFT formalism, and how the mixed anomalies we have described here could be encoded. It is a broader goal to understand the SymTFTs of bosonic $\sigma$-models, for example when the theories of interest arise as symmetry-breaking phases \cite{Antinucci:2024bcm}. An important piece of the puzzle is an understanding of SymTFTs for continuous symmetries, which have been studied recently \cite{Brennan:2024fgj, Antinucci:2024zjp, Bonetti:2024cjk, Gagliano:2024off}. \\

Another structure commonly found in theories relevant to those we have considered here are additional anomalies for these symmetries. For example, if our $\sigma$-model describes a symmetry-breaking phase with some UV completion which itself possesses an anomaly, then anomaly matching requires that the anomaly be reproduced in this low-energy theory. This can result in the addition topological terms to the effective action such as WZW terms, which can have the effect of twisting the magnetic symmetries \cite{Pace:2023mdo}. Anomaly matching may also require the solitons to possess additional protected degrees of freedom such as zero modes. In future work we incorporate these structures into the analysis of the electric-magnetic anomalies and investigate how they contribute. \\

For the symmetries we have considered here, all anomalies are characterized by the inflow of some bulk $(d+1)$-dim anomaly theory which is classified by cohomology $H^{d+2}(BG,\mathbb Z)$. However the modern understanding of anomalies is that they are more generally classified by cobordism classes in $\Omega^{d+2}_{\mathcal X} (BG)$ when the symmetry depends critically on some additional choice of tangential structure $\mathcal X$ \cite{Freed_2021,Kapustin_2015, Yonekura_2019, Gaiotto_2019, Xiong_2018, Freed_2006}. This characterization of the anomaly theories did not make an appearance in the symmetries discussed here, but we anticipate that with the inclusion of the additional structure, such as fermionic degrees of freedom, or the structures discussed above, then this more refined understanding of anomalies will become necessary. It should then be interesting to see how the arguments we have presented here can be generalized to the case when our anomalies are classified by cobordism. \\ 

Finally, the existence of a mixed 't Hooft anomaly provides a convenient method to construct non-invertible symmetries. Indeed, unless the mixed anomaly takes on a particularly simple form, the act of gauging one of the symmetries participating in the anomaly typically yields a non-invertible symmetry in the gauged theory. Thus, the methods shown here provides a new path for the modern endeavor to study and explore novel non-invertible symmetries in quantum field theories. 

\acknowledgments

AS is supported by Simons Foundation award 568420 (Simons Investigator) and award 888994 (The Simons Collaboration on Global Categorical Symmetries). It is my pleasure to thank Ken Intriligator, Daniel Brennan, John McGreevy, Zipei Zhang, Jordan Benson, and Sal Pace for helpful discussions.

\appendix

\section{Appendix: Bundles, Solitons, and the Whitehead Tower}
\label{appa}

We now give some mathematical background, and use it to further elaborate on our claim that magnetic charges (solitons) can be interpreted as defining a connection on a higher $k$-bundle over spacetime, obtained by pulling back the lowest Whitehead tower fibration. 

\subsection{(Higher) Principal Bundles}

Given a group $A$, recall that a \textit{principal bundle} $P$ over a space $M$ is a fiber bundle whose fibers are equivalent to $A$, for which the vertical action along the fibers is just the group action of $A$ on $A$.

\[\begin{tikzcd}
	A && P \\
	\\
	&& M
	\arrow[from=1-1, to=1-3]
	\arrow[from=1-3, to=3-3]
\end{tikzcd}\]

Given a group $A$ and a manifold $M$, there can be topologically distinct principal bundles $P$ over $M$. There exists a classification of all such principal bundles using the \textit{classifying space} $BA$ of $A$, defined as the base of some universal principal $A$ bundle $A \to EA\to BA$ where $EA$ is contractible. The set of topologically-distinct bundles $P$ is in one-to-one correspondence with the homotopy classes of continuous maps $f : M \to BA$. The map $f$ which defines a principal bundle $P$ is called the \textit{classifying map} of $P$, and it defines it by pulling back the universal bundle $P=f^*EA$:

\[\begin{tikzcd}
	A && {P=f^*EA} && EA && A \\
	\\
	&& M && BA
	\arrow[from=1-1, to=1-3]
	\arrow[from=1-3, to=3-3]
	\arrow[from=1-5, to=3-5]
	\arrow[from=1-7, to=1-5]
	\arrow["f", from=3-3, to=3-5]
\end{tikzcd}\]

For our purposes, we are interested in the case where $A$ is a discrete abelian group. Namely, to make contact with \hyperref[sec3]{Section 3}, we want to take it to be $A=\pi_k(X)$. In this case, the principal bundle $P$ simply reduces to a covering space of $M$. The classifying space is just the first Eilenberg-Maclane space $BA=K(A,1)$. If we use the fact that the homotopy class $[f]$ of a map $f : M\to K(A,1)$ defines a cohomology class $[f]\in H^1(M,A)$, then we see that such bundles are topologically classified by cohomology classes in $H^1(M,A)$. \\

Now, in the case of abelian $A$, there exists a generalization of principal bundles to \textit{higher principal bundles} \cite{Costa:2024wks, Baez:2010ya, 2003math......7200B, Hitchin:1999fh, 2012arXiv1207.0248N}. These are sometimes intimidatingly referred to as higher ``categorifications" of principal bundles, but in the discrete case they have a simple description. Geometrically, recall that principal fiber bundles over $M$ are defined by covering $M$ with charts $U_{\alpha}$. Across the codimension-1 chart boundaries $\sim U_{\alpha}\cap U_{\beta}$ between charts, the bundle is then specified by transition functions $\tau_{\alpha\beta} : U_{\alpha}\cap U_{\beta}\to A$. Very roughly, for a higher \textit{principal 2-bundle}, then in addition to keeping track of transition functions across codimension-1 chart boundaries, we also allow for chart boundaries embedded in $U_{\alpha}\cap U_{\beta}$ which can roughly be thought of as being codimension-2, with associated transition functions across these. More precisely, the overlaps $U_{\alpha}\cap U_{\beta}$ support their own principal bundles, which require their own transition functions. Generalizing, a \textit{principal $k$-bundle} keeps track of transition functions all the way down to codimension-$k$ chart boundaries. \\

But for our purposes, we only care about the topological classification of these principal $k$-bundles in the case where $A$ is abelian and discrete. In this case, the classification is almost identical to the classification of ordinary principal bundles. Indeed, the classifying space for a principal $k$-bundle is just the $k^{th}$ Eilenberg-Maclane space $B^kA=K(A,k)$, so that the principal $k$ bundles over a space $M$ are classified by the homotopy classes of classifying maps $ f: M \to B^kA=K(A,k)$. Using the fact that the homotopy class of a map $f : M \to K(A,k)$ defines a cohomology class $[f] \in H^k(M,A)$, we see that \textit{higher principal $k$-bundles are classified by cohomology classes $H^k(M,A)$}! \\

\subsection{The Two Towers: Postnikov and Whitehead}

We now give a brief textbook review of the Postnikov and Whitehead towers of a topological space $X$. The moral behind these structures is to try to decompose all the homotopy data of the space $X$ into only its homotopy groups $\pi_n(X)$. \\

The \textit{Postnikov tower} of a space $X$ is defined as a tower of fibrations, each of the form $K(\pi_n,n)\to X_n \to X_{n-1}$, with the defining feature that $\pi_k(X_n)= \pi_k(X)$ for $k\leq n$ and $\pi_k(X_n)=0$ for $k >n$. 

\[\begin{tikzcd}
	& X \\
	& \vdots \\
	{K(\pi_2,2)} & {X_2} \\
	{K(\pi_1,1)} & {X_1} \\
	& {X_0}
	\arrow[from=1-2, to=2-2]
	\arrow[from=2-2, to=3-2]
	\arrow[from=3-1, to=3-2]
	\arrow[from=3-2, to=4-2]
	\arrow[from=4-1, to=4-2]
	\arrow[from=4-2, to=5-2]
\end{tikzcd}\]

The Postnikov stage $X_n$ can be seen as only preserving the homotopy data $\pi_k(X)$ of $X$ below $k\leq n$, and it kills all higher homotopy. Thus, if we only care about homotopy classes of maps $\sigma : M_d \to X$ where $M$ is a $d$-dim manifold, then it is sufficient to replace $X$ by $X_d$. Each Postnikov fibration $K(\pi_n, n)\to X_n\to X_{n-1}$ is defined by a classifying map $\alpha : X_{n-1}\to K(\pi_n,n+1)$ along with a possible twist. \\

On the other hand, the \textit{Whitehead tower} of $X$ is exactly dual to the Postnikov tower. It is defined as a tower of fibrations \textit{over} the space $X$, each taking the form $K(\pi_n,n-1)\to X^n\to X^{n-1}$, with the defining feature that $\pi_k(X^n)=\pi_k(X)$ for $k > n$ and $\pi_k(X^n)=0$ for $k \leq n$: 

\[\begin{tikzcd}
	& \vdots \\
	{K(\pi_2,1)} & {X^2} \\
	{\pi_1} & {X^1} \\
	& X
	\arrow[from=1-2, to=2-2]
	\arrow[from=2-1, to=2-2]
	\arrow[from=2-2, to=3-2]
	\arrow[from=3-1, to=3-2]
	\arrow[from=3-2, to=4-2]
\end{tikzcd}\]

The stages $X^n$ of the Whitehead tower thus take the opposite approach as those of the Postnikov tower: they kill all homotopy groups below $k \leq n$ and keep all groups above $k > n$. We can recognize $X^1$ as just being the universal cover of $X$, since it is simply connected and preserves all higher homotopy groups. In this sense, the higher Whitehead stages can be considered as the higher-homotopy $n$-connected generalization of the universal cover. \\

In fact the stages $X^n$ of the Whitehead tower are constructed as the homotopy duals to those of the Postnikov tower $X_n$ in the following way. If we define the projection map from $X$ down to the Postnikov stage $p_n : X\to X_n$, then we can define the Whitehead stage $X^n$ as the homotopy fiber of $p_n$. Performing this process iteratively allows us to build up the Whitehead tower. \\

\subsection{Interpretation of Solitons}

Now that we have the appropriate mathematical technology, we can elaborate on our interpretation of the solitonic charges $Q$ for the solitonic symmetry associated to the lowest nontrivial homotopy group $\pi_k(X)$, as discussed in \hyperref[sec32]{Sec 3.2}. Recall that from the definition of a solitonic charge $Q$, we extracted that it defines a map from $k$-cycles $\gamma_k \in H_k(M)$ linking the solitonic defect $A_{d-k-1}$ to values of $\pi_k(X)$, which tells us what value of the winding the $\sigma$-model field $\sigma$ picks up after going around $\gamma_k$. Therefore the defect defines a cohomology class $[Q] \in \text{Hom}(M,\pi_k)\cong H^k(M,\pi_k)$. \\

We claimed that the natural setting in which to understand these defects was in the lowest stage of the Whitehead tower: $B^{k-1}\pi_k \to X^k \to X$. In particular, we want to pull this fibration back to a fibration over spacetime along $\sigma : M\to X$, so that we could interpret $Q$ as defining a $k$-bundle over spacetime. But not every fibration in the Whitehead tower defines a $k$-bundle over $X$, so how exactly can we justify this?\\

The key lies in our assumption that $\pi_k(X)$ is the lowest nontrivial homotopy group of $X$. In particular, this implies that the first nontrivial Postnikov stage $X_k$ is actually equal to the Eilenberg-Maclane space $B^k\pi_k =K(\pi_k,k)$. But recall that the Whitehead stage $X^k$ is defined as the homotopy fiber of the projection map $p_k : X\to X_k=B^k\pi_k$. Therefore the first nontrivial Whitehead stage $X^k$ has a natural $k$-bundle structure, and we are free to pull it back to spacetime along $\sigma : M \to X$. The resulting $k$-bundle over spacetime is defined by the classifying map $\sigma \circ p_k : M\to B^k\pi_k=K(\pi_k,k)$, which defines a cohomology class in $H^k(M,\pi_k)$. But the data in this cohomology class is exactly the same data as in the solitonic charge $[Q]\in H^k(M,\pi_k)$! Thus we identify the topological information in the solitonic charge $Q$ as defining this $k$-bundle structure over spacetime. \\

\bibliographystyle{JHEP}
\bibliography{biblio.bib}

\end{document}